\documentstyle[aps,prb,floats,epsf]{revtex}
\begin{document}

\twocolumn[\hsize\textwidth\columnwidth\hsize\csname
@twocolumnfalse\endcsname

\title{Maximally-localized generalized Wannier functions
for composite energy bands}
\author{Nicola Marzari and David Vanderbilt}
\address{Department of Physics and Astronomy, Rutgers University, 
Piscataway, NJ 08855-0849, USA}

\date{July 10, 1997}

\maketitle
\begin{abstract}

We discuss a method for determining the optimally-localized set of
generalized Wannier functions associated with a set of Bloch bands
in a crystalline solid.  By ``generalized Wannier functions'' we
mean a set of localized orthonormal orbitals spanning the same
space as the specified set of Bloch bands.  Although we minimize a
functional that represents the total spread $\sum_n\;\langle
r^2\rangle_n-\langle{\bf r}\rangle_n^2$ of the Wannier functions
in real space, our method proceeds directly from the Bloch
functions as represented on a mesh of k-points, and carries out
the minimization in a space of unitary matrices $U_{mn}^{(\bf k)}$
describing the rotation among the Bloch bands at each k-point.
The method is thus suitable for use in connection with conventional
electronic-structure codes.  The procedure also returns the total
electric polarization as well as the location of each Wannier
center.  Sample results for Si, GaAs, molecular C$_2$H$_4$ and LiCl 
will be presented.

\end{abstract}


\vskip2pc]

\narrowtext

% -------------------------------------------------------------------

\section{Introduction}
\label{sec:intro}

The study of periodic crystalline solids leads naturally
to a representation for the electronic ground state in terms of
extended Bloch orbitals $\psi_{n\bf k}(\bf r)$,
labeled via their band $n$ and crystal-momentum $\bf k$
quantum numbers.  An alternative representation can be derived
in terms of localized orbitals or Wannier functions $w_n(\bf r-R)$,
that are formally defined via a unitary transformation of the
Bloch orbitals, and are labeled in real space according to the
band $n$ and the lattice vector of the unit cell $\bf R$
to which they belong.\cite{wannier,kohn59,blount,kohn73}

The Wannier representation of the electronic problem
is widely known for its usefulness as a starting point for
various formal developments, such as the semiclassical theory
of electron dynamics or the theory of magnetic interactions
in solids.  But until recently, the {\it practical} importance
of Wannier functions in computational electronic structure theory
has been fairly minimal.  However, this situation is now beginning
to change, in view of two recent developments.  First, there is
a vigorous effort underway on the part of many groups to develop
so-called ``order-N'' or ``linear-scaling'' methods, i.e., methods
for which the computational time for solving for the electronic
ground state scales only as the first power of system size,
\cite{gallireview} instead of the third power typical of
conventional methods based on solving for Bloch states.  Many of
these methods are based on solving directly for localized Wannier
or Wannier-like orbitals that span the occupied subspace,
\cite{parrinello,mgc,odgm,kohn93,stechel,kim,gillan,ordejon} and
thus rely on the localization properties of the Wannier functions.
Second, a modern theory of electric polarization of crystalline
insulators has just recently
emerged;\cite{ksv,vks,resta,om,nunes,resta96} it can be formulated
in terms of a geometric phase in the Bloch representation, or
equivalently, in terms of the locations of the Wannier centers.

The linear-scaling and polarization developments are at the heart
of the motivation for the present work.  However, there is another
motivation that goes back to a theme that has recurred frequently
in the chemistry literature over the last 40 years, namely
the study of ``localized molecular orbitals.''
\cite{boys,foster,edmiston,boysr2,weinstein,leonard}
The idea is to carry out, for a given molecule or cluster,
a unitary transformation from the occupied one-particle Hamiltonian
eigenstates to a set of localized orbitals that correspond more
closely to the chemical (Lewis) view of molecular bond-orbitals.
It seems not to be widely appreciated that these
are the exact analogues, for finite systems, of the Wannier
functions defined for infinite periodic systems.
Various criteria have been introduced for defining the localized
molecular orbitals,\cite{boys,foster,edmiston,boysr2}
two of the most popular being the maximization of the
Coulomb \cite{edmiston} or quadratic \cite{boysr2} self-interactions
of the molecular orbitals.  One of the motivations for such approaches
is the notion that the localized molecular orbitals may form the
basis for an efficient representation of electronic correlations
in many-body approaches, and indeed this ought to be equally true
in the extended, solid-state case.

One major reason why the Wannier functions have seen little
practical use to date in solid-state applications is undoubtedly
their non-uniqueness.  Even in the case of a single isolated band,
it is well known that the Wannier functions $w_n(\bf r)$ are not
unique, due to a phase indeterminacy $e^{i\phi_n(\bf k)}$ in the
Bloch orbitals $\psi_{n{\bf k}}({\bf r})$.  For this case, the
conditions required to obtain a set of maximally localized,
exponentially decaying Wannier functions are
known.\cite{kohn59,onffroy}

In the present work we discuss the determination of the maximally
localized Wannier functions for the case of composite bands.
Now a stronger indeterminacy is present, representable by a free
unitary matrix $U_{mn}^{({\bf k})}$ among the occupied Bloch
orbitals at every wavevector.  We require the choice of a
particular set of $U_{mn}^{({\bf k})}$ according to the criterion
that the sum $\Omega$ of the second moments of the corresponding
Wannier functions be minimized.  (This is the exact analogue of the
criteria of Boys\cite{boysr2} for the molecular-orbital case.)
We show that $\Omega$ can be decomposed into a sum of two
contributions.  The first is invariant with respect
to the $U_{mn}^{({\bf k})}$ and reflects the k-space dispersion
of the band projection operator, while the second reflects the extent
to which the Wannier functions fail to be eigenfunctions of the
band-projected position operators.  We show how this formulation
reduces to previous ones in the case of a single isolated band,
or in one dimension, or for centrosymmetric crystals.

We also describe a numerical algorithm for computing the optimally
localized Wannier functions on a k-space mesh.  The algorithm is
designed to operate in a post-processing mode after a conventional
band-structure calculation, taking as its input the Bloch functions
computed on a mesh of k-points.  (Thus, it is {\it not}
a linear-scaling method.)  We present sample results for
the optimally localized Wannier functions in Si, GaAs, molecular
C$_2$H$_4$, and LiCl.  It should be emphasized that this procedure
generates incidentally a set of Wannier-center positions; these
by themselves can sometimes be very useful for analyzing the
electronic polarization of disordered or distorted insulating
materials.

In this work, we have not considered any further generalizations
of the problem, although several interesting possibilities come
to mind.  For example, one could relax the constraint that the
Wannier functions should be orthonormal to each other (in this
case they should probably not be called ``Wannier functions'').
Such functions would correspond to the ``localized orbitals'' or
``support functions'' appearing in certain linear-scaling
methods\cite{parrinello,stechel,gillan}
and in the chemical-pseudopotential
approach.\cite{anderson,bullett,foulkes}
Alternatively, one could retain the orthonormality requirement,
but ask to find a larger set of functions spanning a space
containing the desired bands as a subspace.  For example, in
Si one could ask for a maximally-localized set of four Wannier-like
functions per atom spanning a space twice as large as, but
containing, the space of the four occupied valence bands.\cite{kohn73}
Again, this is very similar to what is done in certain linear-scaling
methods.\cite{stechel,kim,gillan}  These interesting generalizations
deserve investigation, but have not been pursued here.

The manuscript is organized as follows.  The problem is introduced
in Sec.\ \ref{sec:preliminaries}.  Expressions for the spread
functional, and for its decomposition into gauge-invariant and
gauge-dependent parts, are developed first in real space in
Sec.\ \ref{sec:rspace}.  Section \ref{sec:kspace} then formulates
the corresponding expressions in discrete k-space (that is, on a
mesh of wavevectors).  Special features that arise in one
dimension, or for a single isolated band, or for a crystal with
inversion symmetry, are also discussed there, as is the
steepest-descent minimization algorithm that we use.  Some
discussion and speculation about the asymptotic localization
properties, and the real vs.\ complex nature of the Wannier
functions, appear in Sec.\ \ref{sec:miscprop}.  In
Sec.\ \ref{sec:results} we present test results for Si, GaAs,
C$_2$H$_4$, and LiCl systems.  Finally, in Sec.\ \ref{sec:discuss},
we discuss the significance of the work, emphasizing possible
applications of our approach.  Some details of the real-space,
discrete k-space, and continuous k-space formulations are deferred
to Appendices \ref{app:realmin}, \ref{app:kmesh}, and
\ref{app:geometry}, respectively.  In particular, the relationship
of the present work to the theory of adiabatic quantum phases and
quantum distances is discussed in Appendix \ref{app:geometry}.

% -------------------------------------------------------------------

\section{Preliminaries}
\label{sec:preliminaries}

% -------------------------------------------------------------------

%----------%
\subsection{Isolated and composite bands}
\label{sec:isocomp}
%----------%

We confine ourselves here to the case of an inde\-pendent-particle
Hamiltonian $H=p^2/2m+V(\bf r)$ with a real periodic potential
$V(\bf r)$.  We thus assume the absence of electric and magnetic
fields, and we suppress spin.  The eigenfunctions of $H$ are the
Bloch functions $\psi_{n\bf k}(\bf r)$ labeled by band $n$ and
wavevector $\bf k$.

A Bloch band is said to be {\it isolated} if it does not become
degenerate with any other band anywhere in the Brillouin zone (BZ).
Conversely, a group of bands are said to form a {\it composite group}
if they are connected among themselves by degeneracies, but are
isolated from all lower or higher bands.  For example, in Si the
four valence bands form a composite group, while in GaAs the lowest
valence band is isolated and the higher three form a composite
group.

In the case of isolated bands, it is natural to define
Wannier functions individually for each band.  That is, the
Wannier function for band $n$ (together with its periodic images)
spans the same space as does the isolated Bloch band.
In the case of composite bands, however, it is more natural to
consider a set of $J$ ``generalized Wannier functions'' that
(together with their periodic images) span the same space as the
composite set of $J$ Bloch bands.  That is, the ``generalized
Bloch functions'' $\psi_{n{\bf k}}$ that are connected with the
$n$'th generalized Wannier function will not necessarily be
eigenstates of the Hamiltonian at this $\bf k$, but will be related
to them by a $J\times J$ unitary transformation.

The formulation that follows is designed to apply equally to the
isolated and composite cases.  For the isolated case, $J=1$, and
sums over $n$ can be ignored.  For the composite case, the terms
``Bloch function'' and ``Wannier function'' should be understood
to be meant in the generalized sense discussed above.

It may sometimes be convenient to consider a group of bands as
composite even when some of the members are actually isolated.  For
example, one may wish to consider all of the occupied valence bands
of an insulator
as a composite group.  This is rather natural in connection with
linear-scaling algorithms and the theory of electronic
polarization.  Thus, for GaAs, one may choose to regard all four
valence bands as a composite group.  In this case the Wannier
functions will resemble $\sigma$-bonded pairs of $sp^3$ hybrids,
arguably the most natural choice.  Moreover, the GaAs Wannier functions
defined in this way turn out to be considerably more localized than
those of the top three or bottom valence bands separately.  Again,
the formulation below should be taken to apply equally to this
case, with $n$ running over the $J$ adjacent bands that are being
considered as a composite group.

Finally, the formalism applies equally to any
isolated band or composite group that may exist in a metal or
insulator, regardless of occupation.  However, because the expectation
values of physical operators only depend upon occupied states, one
is usually interested in the case of occupied bands in insulators.

%----------%
\subsection{Definitions}
\label{sec:defs}
%----------%

We denote by $w_n(\bf r-R)$ or $\vert{\bf R}n\rangle$ the Wannier
function in cell $\bf R$ associated with band $n$, given in terms
of the Bloch functions as
\begin{equation}
\vert{\bf R}n\rangle = {V\over(2\pi)^3}\int d{\bf k} \,
e^{-i{\bf k}\cdot{\bf R}}\vert\psi_{n{\bf k}}\rangle \;\;,
\label{eq:wdef}
\end{equation}
so that
\begin{equation}
\vert\psi_{n{\bf k}}\rangle=\sum_{\bf R}
e^{i{\bf k}\cdot{\bf R}}\,\vert{\bf R}n\rangle \;\;.
\label{eq:bdef}
\end{equation}
Here $V$ is the real-space primitive cell volume.  It
is easily shown that the Wannier functions form an orthonormal set.
As usual, the periodic part of the Bloch function is defined as
\begin{equation}
u_{n{\bf k}}({\bf r}) = e^{-i{\bf k}\cdot{\bf r}}
\psi_{n{\bf k}}({\bf r}) \;\;.
\label{eq:udef}
\end{equation}
As shown by Blount,\cite{blount} matrix elements of the position
operator between Wannier functions take the form
\begin{equation}
\langle{\bf R}n\vert{\bf r}\vert{\bf 0}m\rangle = i\,{V\over(2\pi)^3}
\int d{\bf k} \, e^{i{\bf k}\cdot{\bf R}}
\langle u_{n{\bf k}}\vert\nabla_{\bf k}\vert u_{m{\bf k}}\rangle \;\;,
\label{eq:rmatel}
\end{equation}
the converse relation being
\begin{equation}
\langle u_{n{\bf k}}\vert\nabla_{\bf k}\vert u_{m{\bf k}}\rangle
= -i \sum_{\bf R} e^{-i{\bf k}\cdot{\bf R}}
\langle{\bf R}n\vert{\bf r}\vert{\bf 0}m\rangle \;\;.
\label{eq:umatel}
\end{equation}
In equations like these the $\nabla_{\bf k}$ is understood to
act to the right, i.e., only on the ket.
The consistency of these two equations is easily checked; the
latter can be derived by noting that
\begin{eqnarray}
\langle u_{n{\bf k}}\vert u_{m,{\bf k+b}}\rangle &=&
\langle \psi_{n{\bf k}}\vert e^{-i{\bf b\cdot r}}\vert
\psi_{m,{\bf k+b}}\rangle
\nonumber \\
&=& \sum_{\bf R} e^{-i{\bf k\cdot R}}\langle{\bf R}n\vert
 e^{-i{\bf b\cdot r}}\vert {\bf 0}m\rangle \;\;,
\nonumber
\end{eqnarray}
and then equating first orders in $\bf b$.  Similarly, equating second
orders in $\bf b$ leads to
\begin{equation}
\langle{\bf R}n\vert r^2 \vert{\bf 0}m\rangle = -
{V\over(2\pi)^3} \int d{\bf k}\,
e^{i{\bf k}\cdot{\bf R}}
\langle u_{n{\bf k}}\vert\nabla_{\bf k}^2\vert u_{m{\bf k}}\rangle
\;\;.
\label{eq:rrmatel}
\end{equation}
Introducing the notation
$\bar{\bf r}_n=\langle{\bf 0}n\vert {\bf r} \vert{\bf 0}n\rangle$
and
$\langle r^2 \rangle_n=\langle{\bf 0}n\vert r^2 \vert{\bf 0}n\rangle$
for the diagonal elements in the cell at the origin, we have
\begin{equation}
\bar{\bf r}_n = i \, {V\over(2\pi)^3} \int d{\bf k}\,
\langle u_{n{\bf k}}\vert\nabla_{\bf k}\vert u_{n{\bf k}}\rangle
\label{eq:rdiag}
\end{equation}
and
\begin{equation}
\langle r^2 \rangle_n = {V\over(2\pi)^3} \int d{\bf k}\;
\Bigl\vert \vert\nabla_{\bf k} u_{n{\bf k}}\rangle\Bigr\vert^2 \;\;.
\label{eq:rrdiag}
\end{equation}
This last follows from Eq.\ (\ref{eq:rrmatel}) after an integration
by parts.

%----------%
\subsection{Arbitrariness in definition of Wannier functions}
\label{sec:arb}
%----------%

As is well known, Wannier functions are not unique.  For a single
isolated band, the freedom in choice of the Wannier functions
corresponds to the freedom in the choice of the phases of the
Bloch orbitals as a function of wavevector $\bf k$.  Thus, given
one set of Bloch orbitals and associated Wannier functions, another
equally good set is obtained from
\begin{equation}
\vert u_{n{\bf k}}\rangle \;\rightarrow\;
e^{i\phi_n({\bf k})} \vert u_{n{\bf k}}\rangle
\label{eq:phase}
\end{equation}
where $\phi_n$ is a real function of $\bf k$.  Such a transformation
preserves the Wannier center $\bar{\bf r}_n$ modulo a lattice
vector,\cite{blount,ksv,vks} but of course it does not preserve the
spread $\langle r^2 \rangle_n - \bar{\bf r}_n^2$.

For a composite set of bands, the corresponding freedom is
\begin{equation}
\vert u_{n{\bf k}}\rangle \;\rightarrow\;
\sum_m U_{mn}^{(\bf k)}\,\vert u_{m{\bf k}}\rangle
\label{eq:unitary}
\end{equation}
where $U_{mn}$ is a unitary matrix that mixes the bands at
wavevector $\bf k$.  Eq.\ (\ref{eq:phase}) can be regarded as
a special case of Eq.\ (\ref{eq:unitary}) that results when the $U$
are chosen diagonal.  The transformation (\ref{eq:unitary}) does
not preserve the individual Wannier centers, but does preserve the
sum of the Wannier centers, modulo a lattice vector.\cite{ksv}
We shall frequently refer to this freedom as a ``gauge freedom'' and
the transformation (\ref{eq:unitary}) as a ``gauge transformation.''

Our goal is to pick out, from among the many arbitrary choices of
Wannier functions, the particular set that is maximally localized
according to some criterion.  Our choice of criterion is introduced
and justified in the following subsection.  Of course, some
arbitrariness will remain: (i) there will always be an arbitrary
overall phase of each of the $J$ Wannier functions;\cite{expl:phase}
(ii) there is a freedom to permute the $J$ Wannier functions among
themselves; and (iii) there is a freedom to translate any one of
the $J$ Wannier functions by a lattice vector (that is, to decide
which Wannier functions belong to the ``home'' unit cell labeled by
$\bf R=0$).  Aside from these trivial remaining degrees of freedom,
we expect to find a unique set of maximally localized Wannier
functions.

% -------------------------------------------------------------------

\section{Spread functional in real space}
\label{sec:rspace}

% -------------------------------------------------------------------

As a measure of the total delocalization or spread of the Wannier
functions, we introduce the functional
\begin{equation}
\Omega=\sum_n\left[ \langle r^2\rangle_n- \bar{\bf r}_n^{\,2} \right]
\label{eq:om1}
\end{equation}
(recall $\bar{\bf r}_n=\langle{\bf r}\rangle_n$).
Eq.\ (\ref{eq:om1}) is to be minimized with respect to the unitary
transformations $U_{mn}^{(\bf k)}$.
A functional of this form has previously appeared as one possible
definition \cite{boysr2} of the ``localized molecular orbitals''
\cite{boys,foster,edmiston,boysr2,weinstein,leonard}
discussed in the chemistry literature.  Other localization
criteria, such as maximizing the sum of Coulomb self-energies of
the orbitals\cite{edmiston} or the the product
of the separations of the centroids\cite{foster}
have also been suggested.
We focus on the Wannier function obtained by minimizing
Eq.\ (\ref{eq:om1}) for the following reasons.
(i) The Wannier functions so determined correspond precisely to
those considered by previous authors for the isolated-band case in
1D\cite{kohn59,blount,kivelson} and 3D.\cite{blount}
(ii) In the 1D multiband case, the optimally localized Wannier
functions defined by minimizing Eq.\ (\ref{eq:om1}) turn out
to be identical to the
eigenfunctions of the projected position operator
$PxP$,\cite{kivelson,niu} as will be demonstrated shortly.
(Here $P$ is the projection operator onto the group of bands under
consideration,
\begin{equation}
P=\sum_{{\bf R}n} \vert{\bf R}n\rangle\langle{\bf R}n\vert
= \sum_{n{\bf k}}  \vert\psi_{n\bf k}\rangle\langle\psi_{n\bf k}\vert
\;\;,
\label{eq:P}
\end{equation}
and $Q=1-P$ is the projection operator onto all other bands.)
(iii) It is one of the functionals proposed in the chemistry
literature.\cite{boysr2}
(iv) It leads to a particularly elegant formalism, allowing,
for example, the  decomposition into invariant, diagonal, and
off-diagonal contributions as described below.

We find it convenient to decompose the functional (\ref{eq:om1})
into two terms,
\begin{equation}
\Omega=\Omega_{\rm\,I}+\widetilde{\Omega} \;\;,
\label{eq:om2}
\end{equation}
where
\begin{equation}
\Omega_{\rm\,I} = \sum_n \left[ \langle r^2\rangle_n- \sum_{{\bf R}m}
\,\Bigl\vert \langle{{\bf R}m}\vert{\bf r}\vert
{\bf 0}n\rangle\Bigr\vert^2 \right]
\label{eq:om0}
\end{equation}
and
\begin{equation}
\widetilde\Omega = \sum_n \sum_{{\bf R}m\ne{\bf 0}n}
\Bigl\vert \langle{{\bf R}m}\vert{\bf r}\vert
{\bf 0}n\rangle\Bigr\vert^2 \;\;.
\label{eq:omt}
\end{equation}
Clearly the second term is positive definite.  While it is not
immediately obvious, the first term is also positive definite,
and moreover it is {\it gauge-invariant} (i.e., independent
of the choice of unitary transformations among the bands).  To see
this, we use the definitions of $P$ and $Q$ in terms of the Wannier
functions to write
\begin{eqnarray}
\Omega_{\rm\,I} &=& \sum_{n\alpha}
\langle{{\bf 0}n}\vert r_\alpha Q r_\alpha \vert{\bf 0}n\rangle
\nonumber \\
&=& \sum_\alpha {\rm tr}_{\rm c}[P r_\alpha Q r_\alpha]
\nonumber \\
&=&
\Vert PxQ \Vert_{\rm c}^2 +
\Vert PyQ \Vert_{\rm c}^2 +
\Vert PzQ \Vert_{\rm c}^2
\;\;.
\label{eq:trpc}
\end{eqnarray}
Here $\rm tr_c$ indicates the trace per unit cell, and
$\Vert A \Vert_{\rm c}^2 = {\rm tr}_{\rm c}[A^\dagger A]$.
The last form makes it obvious that $\Omega_{\rm\,I}$ is positive
definite.  Operators of the form $P{\bf r}Q$ have been discussed
extensively by Nenciu;\cite{nenciuPxQ} unlike {\bf r} itself,
$P{\bf r}Q$ commutes with lattice translations, and its expectation
value is well defined in any (normalizable) extended state.
Thus, it follows that $\Omega_{\rm\,I}$ is
gauge-invariant (i.e., invariant with respect to the choice of
Wannier functions, or equivalently to the choice of the unitary
mixing matrices $U_{mn}^{\bf(k)})$.  This will become even clearer
in Sec.\ \ref{sec:kspace}, where $\Omega_{\rm\,I}$ is expressed
in a finite-difference k-space representation.

It was stated earlier that in 1D the set of Wannier functions that
minimizes the spread functional, Eq.\ (\ref{eq:om1}), turns out to
be identical to the set of eigenfunctions of the projected position
operator $PxP$.  This can now be seen as follows.  Choose the
Wannier functions $\vert 0m\rangle$ to be eigenfunctions of $PxP$
with associated eigenvalues $\bar{x}_{0m}$.  Then
\begin{equation}
\langle Rn\vert x \vert 0m\rangle =
\langle Rn\vert PxP \vert 0m\rangle =
\bar{x}_{0m} \, \delta_{R,0} \, \delta_{m,n} \;\;.
\label{eq:eigfun}
\end{equation}
Clearly $\widetilde{\Omega}$ vanishes, and since $\Omega_{\rm I}$
is gauge-invariant, this minimizes Eq.\ (\ref{eq:om2}).  Thus in
1D the solution is essentially trivial, even in the multi-band
case, and $\Omega_{\rm min}=\Omega_{\rm\,I}$ at the solution.

From this point of view, it can now be understood that the
essential difficulty in the 3D case is that the operators $PxP$,
$PyP$, and $PzP$ do not commute (or, in the language of Appendix
\ref{app:realmin}, that matrices $X$, $Y$, and $Z$ do not commute.)
For if they did, one could choose the Wannier functions to be
simultaneous eigenfunctions of all three, and one could again make
$\widetilde\Omega$ vanish.  But this is not generally the case, and
the problem is to find a set of Wannier functions that makes the
best possible compromise in the attempt to diagonalize all three
simultaneously.  Indeed, it appears very natural that the criterion
should be simply to reduce, as far as possible, the mean-square
average of all off-diagonal matrix elements of $x$, $y$, and $z$
between Wannier functions; this is precisely the criterion encoded
into $\widetilde\Omega$.  A procedure for carrying out this
minimization directly in real space is sketched in Appendix
\ref{app:realmin}.  However, for crystalline solids with periodic
boundary conditions, it is more straightforward to work in k-space
as discussed in the following section.

Finally, for later reference, it is useful to decompose
$\widetilde\Omega$ into band-off-diagonal and band-diagonal
pieces,
\begin{equation}
\widetilde\Omega=\Omega_{\rm OD} + \Omega_{\rm D} \;\;,
\label{eq:dod}
\end{equation}
where
\begin{equation}
\Omega_{\rm OD} = \sum_{m\ne n} \sum_{\bf R}
\,\Bigl\vert \langle{{\bf R}m}\vert{\bf r}\vert
{\bf 0}n\rangle\Bigr\vert^2 \;\;.
\label{eq:omod}
\end{equation}
and
\begin{equation}
\Omega_{\rm D} = \sum_n \sum_{\bf R\ne0}
\,\Bigl\vert \langle{{\bf R}n}\vert{\bf r}\vert
{\bf 0}n\rangle\Bigr\vert^2 \;\;.
\label{eq:omd}
\end{equation}
%

% -------------------------------------------------------------------

\section{Spread functional in k-space}
\label{sec:kspace}

% -------------------------------------------------------------------

%----------%
\subsection{Transition to k-space}
\label{sec:ktransit}
%----------%

We now derive expressions for $\Omega$, $\Omega_{\rm I}$,
$\widetilde\Omega$, etc.\ in terms of a discretized k-space mesh.
We begin by substituting expressions (\ref{eq:rdiag}) and
(\ref{eq:rrdiag}) into Eq.\ (\ref{eq:om1}), and making use of
\begin{equation}
{V\over(2\pi)^3}\int d{\bf k} \;\rightarrow\; {1\over N} \sum_{\bf k}
\;\;,
\nonumber
\end{equation}
where $N$ is the number of real-space cells in the system, or
equivalently, the number of k-points in the Brillouin zone.
Using the finite-difference expressions for $\nabla_{\bf k}$
and $\nabla^2_{\bf k}$ introduced in Appendix \ref{app:kmesh},
we have
\begin{equation}
\bar{\bf r}_n =
{i\over N} \sum_{\bf k,b} w_b \, {\bf b} \,
\left[ \langle u_{n{\bf k}}\vert u_{n,{\bf k+b}}\rangle -1 \right]
\label{eq:rtmp}
\end{equation}
and
\begin{equation}
\langle r^2 \rangle_n =
{1\over N} \sum_{\bf k,b} w_b \, 
\left[ 2-2\,{\rm Re}\,\langle u_{n{\bf k}}
\vert u_{n,{\bf k+b}}\rangle \right] \;\;.
\label{eq:rrtmp}
\end{equation}
Here $\bf b$ are vectors connecting each k-point to its near neighbors
and $w_b$ are associated weights (see Appendix \ref{app:kmesh}).

Clearly, these expressions reduce to Eqs. (\ref{eq:rdiag}) and
(\ref{eq:rrdiag}) in the limit of dense mesh spacing ($N\rightarrow
\infty$, $b\rightarrow0$).  However, we should like to insist on
a second desirable property as well: namely, that for a given
k-mesh, $\bar{\bf r}_n$ and $\langle r^2 \rangle_n$ should
transform as expected when the definition of $\vert{\bf
0}n\rangle$ is shifted by a lattice vector.  (This corresponds to
changing the choice of which Wannier functions belong to the
``home'' unit cell.)  That is, when
$\vert u_{n\bf k}\rangle \rightarrow \vert u_{n\bf k}\rangle
e^{-i\bf k\cdot R}$, so that
$\langle u_{n\bf k} \vert u_{n,\bf k+b}\rangle \rightarrow
\langle u_{n\bf k} \vert u_{n,\bf k+b}\rangle
e^{-i\bf b\cdot R}$, we should find
\begin{eqnarray}
\bar{\bf r}_n &\rightarrow& \bar{\bf r}_n + {\bf R}
\;\;,
\nonumber \\
\langle r^2 \rangle_n &\rightarrow& \langle r^2 \rangle_n
+2\bar{\bf r}_n \cdot {\bf R} + R^2
\;\;,
\label{eq:Rcondit}
\end{eqnarray}
so that $\Omega$ will be unchanged.  Expressions (\ref{eq:rtmp}) and
(\ref{eq:rrtmp}) do not obey these requirements, but can be
modified to do so.  As long as the modifications leave the summands
unchanged to order $b$ and $b^2$ in Eqs.\ (\ref{eq:rtmp}) and
(\ref{eq:rrtmp}) respectively, they will still reduce to 
Eqs.\ (\ref{eq:rdiag}) and (\ref{eq:rrdiag}) in the continuum limit.

Let
\begin{equation}
M_{mn}^{(\bf k,b)} = \langle u_{m\bf k}\vert u_{n,\bf k+b}\rangle
\label{eq:mdef}
\end{equation}
and, for a given $n$, $\bf k$, and $\hat{\bf b}$, let
\begin{equation}
M_{nn}^{(\bf k,b)} = 1+ixb+{1\over 2}yb^2+{\cal O}(b^3) \;\;.
\nonumber
\end{equation}
By expanding $\langle u_{n,\bf k+b}\vert u_{n,\bf k+b}\rangle=1$
order by order in $b$, it is easy to check that $x$ and $y$ are
real.  Then, referring to Eqs.\ (\ref{eq:rtmp}) and
(\ref{eq:rrtmp}), we have
\begin{equation}
M_{nn}^{(\bf k,b)} -1 = ixb+{\cal O}(b^2) \;\;,
\label{eq:rep1}
\end{equation}
\begin{equation}
2-2\,{\rm Re}\,M_{nn}^{(\bf k,b)} = -yb^2+{\cal O}(b^3) \;\;.
\label{eq:rep2}
\end{equation}
It is also easy to check that
\begin{equation}
ixb=i\,{\rm Im}\,\ln M_{nn}^{(\bf k,b)} +{\cal O}(b^2) \;\;,
\label{eq:sub1}
\end{equation}
\begin{equation}
-yb^2=1-\vert M_{nn}^{(\bf k,b)}\vert^2 +x^2b^2 +{\cal O}(b^3) \;\;.
\label{eq:sub2}
\end{equation}
Thus, in place of Eq.\ (\ref{eq:rtmp}) we write
\begin{equation}
\bar{\bf r}_n = - {1\over N} \sum_{\bf k,b} w_b \, {\bf b} \,
{\rm Im}\,\ln M_{nn}^{(\bf k,b)} \;\;,
\label{eq:r}
\end{equation}
and, in place of Eq.\ (\ref{eq:rrtmp}),
\begin{equation}
\langle r^2 \rangle_n = {1\over N} \sum_{\bf k,b} w_b \,
\left\{
  \left[
     1-\vert M_{nn}^{(\bf k,b)}\vert^2
  \right] +
  \left[
     {\rm Im}\,\ln M_{nn}^{(\bf k,b)}
  \right]^2
\right\} \;\;.
\label{eq:rr}
\end{equation}
When inserted in Eq.\ (\ref{eq:om1}), this gives our operational
definition of the spread functional $\Omega$.

It is easy to check that Eqs.\ (\ref{eq:r})-(\ref{eq:rr}) obey
conditions (\ref{eq:Rcondit}) exactly, while still reducing to
Eqs.\ (\ref{eq:rdiag}) and (\ref{eq:rrdiag}) in the continuum
limit.  The expression for the Wannier center, Eq.\ (\ref{eq:r}),
is strongly reminiscent of the Berry-phase expression of
Refs.\onlinecite{ksv} and
\onlinecite{vks}, and reduces to it for an isolated band
in 1D.  [It is also exactly invariant, modulo a lattice vector,
under any change of phases of the form of Eq.\ (\ref{eq:phase}),
provided that the phases still vary smoothly enough with $\bf k$
to prevent ambiguity in the choice of branch when evaluating $\ln
M_{nn}^{(\bf k,b)}$.  Of course, it is not invariant under an
arbitrary gauge transformation, Eq.\ (\ref{eq:unitary})].

Note that expression (\ref{eq:rr}) for $\langle r^2 \rangle_n$ is
not unique, even when insisting on the invariance condition
(\ref{eq:Rcondit}).  For example, replacing
\begin{equation}
1-\vert M_{nn}^{(\bf k,b)}\vert^2
\;\rightarrow\;
-2\,{\rm Re}\, \ln M_{nn}^{(\bf k,b)}
\label{eq:replace}
\end{equation}
results in an equally valid
finite-difference formula for $\Omega$.  However, use of the form
(\ref{eq:rr}) facilitates a connection with the decomposition
of $\Omega=\Omega_{\rm I}+\Omega_{\rm OD}+\Omega_{\rm D}$ into
invariant, off-diagonal, and diagonal components as in
Eqs.\ (\ref{eq:om2}) and (\ref{eq:dod}).  Following the lines of
the formalism above, one finds that Eq.\ (\ref{eq:om0}) becomes
\begin{eqnarray}
\Omega_{\rm I} &=& {1\over N} \sum_{\bf k,b} w_b \,
\left( J-\sum_{mn} \vert M_{mn}^{(\bf k,b)} \vert^2 \right)
\nonumber \\
               &=& {1\over N} \sum_{\bf k,b} w_b \,
{\rm tr}\,[P^{\bf(k)}Q^{\bf(k+b)}] \;\;,
\label{eq:omI}
\end{eqnarray}
where $P^{\bf(k)}=\sum_n \vert u_{n\bf k}\rangle
\langle u_{n\bf k}\vert$, $Q^{\bf(k)}=1-P^{\bf(k)}$, and the band
indices $m,n$ run over $1,...,J$.  Similarly, Eqs.\ (\ref{eq:omod})
and (\ref{eq:omd}) become
\begin{equation}
\Omega_{\rm OD}= {1\over N} \sum_{\bf k,b} w_b
\sum_{m\ne n} \vert M_{mn}^{\bf (k,b)} \vert^2
\label{eq:omod2}
\end{equation}
and
\begin{equation}
\Omega_{\rm D}= {1\over N} \sum_{\bf k,b} w_b
\sum_n \left( - {\rm Im}\,\ln M_{nn}^{\bf (k,b)} 
 - {\bf b}\cdot\bar{\bf r}_n \right)^2 \;\;.
\label{eq:omd2}
\end{equation}
From these expressions, it is again evident that $\Omega_{\rm I}$,
$\Omega_{\rm OD}$, and $\Omega_{\rm D}$ are all positive definite.

Eq.\ (\ref{eq:omI}) also now shows clearly that $\Omega_{\rm I}$
is gauge-invariant [i.e., independent of the choice of the Wannier
functions, Eq.\ (\ref{eq:unitary})].  Heuristically, $\Omega_{\rm I}$
represents the degree of dispersion of the band projection operator
$P^{\bf (k)}$ through the Brillouin zone.  That is, $\Omega_{\rm I}$
is small insofar as $P^{\bf (k)}$ is nearly independent of $\bf k$.
(Note that ${\rm tr}[P_1 Q_2]=\Vert P_1-P_2\Vert^2/2$ represents the
``spillage,'' or degree of mismatch, between the spaces 1 and 2.)
Since $\Omega_{\rm I}$ is invariant with respect to
gauge transformations (\ref{eq:unitary}), it can be evaluated
once and for all in the initial gauge (i.e., using the initial
$u_{n\bf k}$) before performing the minimization procedure outlined
below.

It is amusing to note, following the ideas of
Refs.\ \onlinecite{pati,anandan,joshi}, that one can define a
``quantum distance'' between two wavevectors $\bf k$ and $\bf k'$
as $dl^2={\rm tr}[P^{\bf(k)}Q^{\bf(k')}]$, thus inducing a metric
upon the k-space.  The invariant part of the
spread functional, $\Omega_{\rm I}$, turns out to be nothing
other than the Brillouin-zone average of the trace of this metric.
We discuss the properties of this metric, and speculate about
its utility, in Appendix \ref{app:geometry}.

%----------%
\subsection{Gradient of spread functional}
\label{sec:grad}
%----------%

We now consider the first-order change of the spread functional
$\Omega$ arising from an infinitesimal gauge transformation,
Eq.\ (\ref{eq:unitary}), given by
\begin{equation}
U_{mn}^{\bf(k)}=\delta_{mn}+dW_{mn}^{\bf(k)} \;\;,
\label{eq:rot}
\end{equation}
where $dW$ is an infinitesimal antiunitary matrix,
$dW^\dagger=-dW$, so that
\begin{equation}
\vert u_{n\bf k}\rangle \;\rightarrow\; \vert u_{n\bf k}\rangle
+ \sum_m dW_{mn}^{\bf(k)} \, \vert u_{m\bf k}\rangle \;\;.
\label{eq:newu}
\end{equation}
We seek an expression for $d\Omega/dW_{mn}^{\bf(k)}$.  We use
the convention
\begin{equation}
\left({dF\over dW}\right)_{nm}= {dF\over dW_{mn}}
\label{eq:dfdw}
\end{equation}
(note the reversal of indices), so that
\begin{equation}
{d\,{\rm tr}\,[dW\,B] \over dW} = B \;\;,
\label{eq:trdw}
\end{equation}
\begin{equation}
{d\,{\rm Re\,tr}\,[dW\,B] \over dW} = {\cal A}[B] \;\;,
\label{eq:trredw}
\end{equation}
\begin{equation}
{d\,{\rm Im\,tr}\,[dW\,B] \over dW} = {\cal S}[B] \;\;,
\label{eq:trimdw}
\end{equation}
where $\cal A$ and $\cal S$ are the superoperators
${\cal A}[B]=(B-B^\dagger)/2$ and ${\cal S}[B]=(B+B^\dagger)/2i$.
As we shall see shortly, it is possible to cast $d\Omega$ into
the form of the numerators of Eqs.\ (\ref{eq:trredw})
and (\ref{eq:trimdw}).

For the present purpose it is convenient to write
$\Omega=\Omega_{\rm I,OD}+\Omega_{\rm D}$, where
$\Omega_{\rm D}$ is the diagonal part given by
Eq.\ (\ref{eq:omd2}), and the invariant and off-diagonal
parts are combined into
\begin{eqnarray}
\Omega_{\rm I,OD} &=& \Omega_{\rm I} + \Omega_{\rm OD} \nonumber \\
&=&  {1\over N} \sum_{\bf k,b} w_b \,
\sum_n \left[\,1\,-\,\vert M_{nn}^{\bf(k,b)} \vert^2\,\right] \;\;.
\label{eq:iod}
\end{eqnarray}
From Eq.\ (\ref{eq:newu}) it follows that
\begin{equation}
dM_{nn}^{\bf(k,b)} = -[ dW^{\bf(k)} M^{\bf(k,b)} ]_{nn}
                   +[ M^{\bf(k,b)} dW^{\bf(k+b)} ]_{nn} \;\;.
\label{eq:dM}
\end{equation}
Using $M^{\bf(k,b)}=[M^{\bf(k+b,-b)}]^\dagger$ and $dW=-dW^\dagger$,
the second term in Eq.\ (\ref{eq:dM}) can be transformed to become
$-[dW^{\bf(k+b)}M^{\bf(k+b,-b)}]_{nn}^*$.  Defining
\begin{equation}
R_{mn}^{\bf(k,b)} = M_{mn}^{\bf(k,b)} M_{nn}^{\bf(k,b)*} \;\;,
\label{eq:Rdef}
\end{equation}
we thus find
\begin{equation}
d\Omega_{\rm I,OD} =  {4\over N} \sum_{\bf k,b} w_b \,
{\rm Re\,tr}\,[ dW^{\bf(k)} R_{mn}^{\bf(k,b)} ] \;\;.
\label{eq:diod}
\end{equation}
Similarly, defining
\begin{equation}
q_n^{\bf(k,b)} = {\rm Im}\,\ln M_{nn}^{\bf(k,b)}
+ {\bf b\cdot\bar{r}}_n
\label{eq:qdef}
\end{equation}
and
\begin{equation}
\widetilde{R}_{mn}^{\bf(k,b)} = {M_{mn}^{\bf(k,b)}\over
M_{nn}^{\bf(k,b)}} \;\;,
\label{eq:Rtdef}
\end{equation}
Eq.\ (\ref{eq:omd2}) gives for the diagonal part
\begin{eqnarray}
d\Omega_{\bf D} = {2\over N} \sum_{\bf k,b} w_b \sum_n
&& q_n^{\bf(k,b)}\,{\rm Im}\,[
   -dW^{\bf(k)} \widetilde{R}^{\bf(k,b)}
\nonumber \\
   &&+dW^{\bf(k+b)} \widetilde{R}^{\bf(k+b,-b)}
]_{nn} \;\;.
\label{eq:long}
\end{eqnarray}
Substituting $q_n^{\bf(k+b,-b)}=-q_n^{\bf(k,b)}$, the two terms can
be combined, resulting in
\begin{equation}
d\Omega_{\bf D} = - {4\over N} \sum_{\bf k,b} w_b \,
{\rm Im\,tr}\, [ dW^{\bf(k)} T^{\bf(k,b)} ] \;\;,
\label{eq:long2}
\end{equation}
where
\begin{equation}
T_{mn}^{\bf(k,b)} = \widetilde{R}_{mn}^{\bf(k,b)} q_n^{\bf(k,b)}
\;\;.
\label{eq:Tdef}
\end{equation}
We thus arrive at the desired expression for the gradient
of the spread functional,
\begin{equation}
G^{\bf(k)} = {d\Omega\over dW^{\bf(k)}} = 4 \sum_{\bf b} w_b
\left( {\cal A}[R^{\bf(k,b)}] - {\cal S}[T^{\bf(k,b)}]
\right) \;\;.
\label{eq:grado}
\end{equation}
We note, for completeness, that making the replacement
(\ref{eq:replace}) has just the effect of replacing $R$ by
$\widetilde{R}$ in the first term above.

The condition for having found a minimum is that the above
expression should vanish.  We discuss the numerical minimization
of the spread functional by steepest descents, using this
gradient expression, in Sec.\ \ref{sec:sd}.

%----------%
\subsection{Special cases}
\label{sec:cases}
%----------%

\subsubsection{One dimension}
\label{sec:1D}

As mentioned in Sec.\ \ref{sec:rspace}, in 1D it should be possible
to choose the Wannier functions to be eigenfunctions of
the band-projected position operator $PxP$, and thus to make
$\widetilde{\Omega}=\Omega_{\rm OD}+\Omega_{\rm D}$ vanish.
Unfortunately, on a finite k-mesh $\widetilde{\Omega}$ cannot
generally be made to vanish completely.  At the minimum,
$\Omega_{\rm D}$ does vanish, but $\Omega_{\rm OD}$ does not,
leaving a remainder that is expected to approach zero as
${\cal O}(b^2)$ with mesh spacing $b$.

First, note that starting from any given gauge, it is
straightforward to adjust the phases of the $\vert u_{nk_j}\rangle$
in order to make $\Omega_{\rm D}=0$ without affecting
$\Omega_{\rm OD}$ whatsoever.  For each $n$, let
$\lambda_n=s_n/|s_n|$ where $s_n=\prod_{j=0}^{N-1}
M_{nn}^{(k_j,+b)}$ (thus $\lambda_n$ is the ``Berry phase'' of band
$n$); then, starting from the first point $j=0$, recursively set
the phase of $\vert u_{n,k_j+b}\rangle$ such that
$M_{nn}^{(k_j,+b)}=\lambda_n^{1/N}$, for successive k-points $j$.
Then all the $M_{nn}^{(k_j,+b)}$ will have the same phase and
$\Omega_{\rm D}$ will vanish.  This operation has no effect
whatsoever on the magnitudes of the elements of $M_{mn}^{(k,+b)}$,
and so, by Eq.\ (\ref{eq:omod2}), it leaves $\Omega_{\rm OD}$
unchanged.  This argument demonstrates that $\Omega_{\rm D}=0$ and
thus $\widetilde\Omega=\Omega_{\rm OD}$ at the minimum.

A good starting guess that will make $\Omega_{\rm OD}$ rather
small (and keep $\Omega_{\rm D}=0$) can be constructed as follows.
We first establish a notion of ``parallel transport'' of the
Bloch functions.  Starting with some arbitrary choice (from
among all possible $J\times J$ unitary rotations) of the
$\vert u_{nk_0}\rangle$ at an initial k-point $k_0$, we choose
the $\vert u_{n,k_0+b}\rangle$ at the next point $k_0+b$ by
insisting that $M_{mn}^{(k_0,+b)}$ should be hermitian.  [This choice
is uniquely given by the singular value decomposition
$M=V\Sigma W^\dagger$, where $V$ and $W$ are unitary and
$\Sigma$ is a diagonal matrix with nonnegative diagonal elements.
Then $M=(V\Sigma V^\dagger)(VW^\dagger)$; and by appropriate
unitary rotation, the $VW^\dagger$ term can be eliminated,
leaving $M$ hermitian.]  This procedure is repeated, progressing
from k-point to k-point [and using $u_{n,-\pi/a}(x)=
u_{n,\pi/a}(x)\,e^{2\pi ix/a}$ when crossing the Brillouin
zone boundary] until the loop is completed, establishing
a new set of states at $k_0$ that are related to the
initial ones by a unitary transformation $\Lambda$.
(This matrix $\Lambda$ is the generalization of the Berry
phase\cite{berry} to a non-Abelian multi-dimensional
manifold.\cite{resta96,wilczek,mead,changniu})  Next, one diagonalizes
$\Lambda=V\lambda V^\dagger$, and rotates the bands at
every k-point by the same unitary matrix $V$.  Having
done this, one finds that each state $\vert u_{nk_0}\rangle$
is carried onto itself by parallel transport around
the loop, except that it returns with an excess phase
$\lambda_n$.  Finally, defining $\gamma_n=\lambda_n^{-1/N}$
and modifying the phases as $\vert u_{nk_j}\rangle \rightarrow
\gamma_n^j \, \vert u_{nk_j}\rangle$, we arrive at the
desired solution (parallel-transport gauge).

At this solution, each $M_{mn}^{(k,+b)}= K_{mn}^{(k)} \gamma_n$
with $K$ hermitian.  It follows that the
${\rm Im}\,\ln M_{nn}^{(k,+b)}$ are independent of $k$,
the Wannier centers $\bar{x}_n$ are determined by the
$\lambda_n$, and thus $\Omega_{\rm D}$ of Eq.\ (\ref{eq:omd2})
vanishes.  From Eq.\ (\ref{eq:omod2}) it can be seen that
$\Omega_{\rm OD}$ does not generally vanish.  However,
$\Omega_{\rm OD}$ depends only on the matrices $K_{mn}^{(k)}$,
and these can be shown to scale as $\delta_{mn}+{\cal O}(b^2)$,
so that $|M_{mn}^{(k,+b)}|^2$ is expected to scale as
${\cal O}(b^4)$, and $\Omega_{\rm OD}$ as ${\cal O}(b^2)$.

If a minimization of $\Omega$ is then carried out starting from
this parallel-transport solution, one expects $\Omega_{\rm D}$ to
remain zero and $\Omega_{\rm OD}$ to be reduced slightly, the
reduction again being expected to be ${\cal O}(b^2)$.
The Wannier centers will also presumably shift slightly.

If one is mainly interested in the Wannier centers in the 1D
case, it may be preferable to take these from the parallel-transport
solution (i.e., from the $\lambda_n$), rather than from the
$\bar{x}_n$ at the minimum.  The former approach corresponds more
closely with the Berry-phase viewpoint,\cite{ksv,vks,resta,resta96}
and indeed the sum of the Wannier centers so defined corresponds
to the usual formula for the electronic polarization.\cite{ksv,vks}
(Actually, for this purpose, the full parallel-transport construction
need not be carried out.  $\Lambda$ may be calculated
as the product of the unitary parts of the $M$ matrices in
any given representation, and the $\lambda_n$ obtained as its
eigenvalues.  By ``unitary part'' we mean the $VW^\dagger$ taken
from the singular value decomposition $M=V\Sigma W^\dagger$.)
On the other hand, the parallel-transport
formulation does not easily generalize to higher dimensions.
Thus, the approach of minimizing the $\Omega$ functional appears
to be the most natural one in higher dimensions, and it
gives results that differ only very slightly from the
parallel-transport solution for reasonable meshes in 1D.

\subsubsection{Isolated band in multiple dimensions}
\label{sec:isoband}

For the case of an isolated band in multiple dimensions, the
problem of finding the optimally localized Wannier function
maps onto the problem of solving the Laplace equation for a
phase field,\cite{blount,weinreich} as described next.
$\Omega_{\rm OD}$ is not present,
and the problem reduces to minimizing $\Omega_{\rm D}$, so that
only the second term in Eq.\ (\ref{eq:grado}) appears.
Clearly $\widetilde{R}$ is identically one and
$T^{\bf(k,b)}=q^{\bf(k,b)}$ is real, so that Eq.\ (\ref{eq:grado})
becomes
\begin{equation}
G^{\bf(k)}=4i \sum_{\bf b} w_b \, {\rm Im}\,\ln M^{\bf(k,b)} \;\;.
\label{eq:grado1}
\end{equation}
At the solution, this expression must vanish.  Starting from
some initial guess on the phases of the $\vert u_{\bf k}\rangle$
and making the substitution of Eq.\ (\ref{eq:phase}), it can be
seen that Eq.\ (\ref{eq:grado1}) corresponds to a solution of the
Laplace equation for the phase field $\phi({\bf k})$.  This
corresponds closely to the discussion in the vicinity of Eq.\ (5.15)
of Ref.\ \onlinecite{blount}.

The quantity $-\sum_b w_b\,{\bf b}\,{\rm Im}\,\ln M^{\bf(k,b)}$ is a
finite-difference representation of the vector field
${\bf A}({\bf k})=i\langle u_{\bf k}\vert\nabla_{\bf k}\vert u_{\bf k}
\rangle$; in the language of the theory of geometrical phases,
$\bf A(k)$ is known as the ``gauge potential'' or
``Berry connection.''\cite{resta96,berry,mead}
The average value of $\bf A(k)$ is gauge-invariant (modulo a
quantum) and is set by the Berry phase,\cite{ksv,vks,resta}
but $\bf A(k)$ is locally gauge-dependent.
The minimization of $\Omega$ via the solution of the Laplace
equation selects the gauge that makes $\nabla\cdot\bf A$ vanish,
but its curl, ${\bf B}=\nabla\times\bf A$, is generally non-zero.  In
fact, $\bf B$, which is known as the ``Berry curvature,'' is a
gauge-invariant quantity; it can be regarded as an intrinsic
property of the band.\cite{resta96,changniu}

Since $\bf A(k)$ is periodic in $k$-space, one can alternatively
think in terms of the Fourier coefficients $\bf A(R)$.  These can
be divided into three contributions: the {\it uniform} part,
$\bf A(R\!=\!0)$; and, for ${\bf R}\ne{\bf 0}$, the {\it longitudinal}
and {\it transverse} parts $A_{\rm L}({\bf R})$ and
${\bf A}_{\rm T}({\bf R})$, i.e., the components of $\bf A(R)$
parallel and perpendicular to $\hat{\bf R}$, respectively.
The uniform part gives the Wannier center; the longitudinal part
is the part that can be made to vanish by appropriate choice of
gauge; and the transverse part is gauge-invariant (it is related
to the Berry curvature) and determines
the minimum value of  $\Omega_{\rm D}$.  In fact, the individual
Fourier components $\bf A(R)$ can be related to the matrix
elements $\langle{\bf R}|{\bf r}|{\bf 0}\rangle$ of
Eq.\ (\ref{eq:omt}); it thus follows that at the solution, the
latter are purely transverse, ${\bf A}({\bf R})\cdot {\bf R}=0$.
Unfortunately, the picture does not appear to remain so simple
in the multiband case, as discussed in Appendix \ref{app:geometry}.

The Berry curvature, or equivalently, the transverse part of the Berry
connection, can easily be shown to vanish for an isolated
band in a crystal with inversion symmetry (see
Sec.\ \ref{sec:inversion}); in this case the solution for
$\bf A(k)$ is a perfectly uniform one, and $\Omega_{\rm D}$
vanishes at the solution.  In a non-centrosymmetric crystal, however,
this is not the case, since a non-zero Berry curvature is generally
present.  This provides a complementary viewpoint, for
the single-band case, on the fact that the non-invariant part
$\widetilde{\Omega}$ of the spread functional cannot generally be
made to vanish.

\subsubsection{Inversion symmetry}
\label{sec:inversion}

When inversion symmetry $V({\bf r})=V(-{\bf r})$ is present, the
cell-periodic Bloch functions can be chosen to be real in the
reciprocal representation; that is, $u_{n\bf k}({\bf r}) =
\sum_{\bf G}u_{n\bf k}({\bf G}) \exp(i\bf G\cdot r)$ with
$u_{n\bf k}(\bf G)$ real.  It might naively appear that all the
$M_{mn}^{\bf(k,b)}$ matrices could then be chosen real, and that
the solution of the minimization problem might be trivial in some
sense.  This is not quite true.  Even for an isolated band, there
is the complication that the Berry phase of the band may be $-1$
instead of $+1$; in this case the $u_{n\bf k}(\bf G)$ can be chosen
real {\it locally} (i.e., in a small neighborhood around any given
$\bf k$), but not {\it globally}.  But this really only means
that the corresponding Wannier function has definite symmetry
under inversion through a symmetry center (``Wyckoff position'')
other than the one at the origin, and the Berry phase can be
reset to $+1$ by a shift of origin.  For the case of composite bands,
however, the problem is to choose a particular gauge transformation
[Eq.\ (\ref{eq:unitary})], not just a phase transformation
[Eq.\ (\ref{eq:phase})], and for this the presence of inversion
symmetry does not provide any obvious solution.

For example, consider the case of the four valence bands of Si.
(Numerical results for this case appear in Sec.\ \ref{sec:si}.)
Taking the origin at the center of the bond oriented along $[111]$,
it turns out to be possible to choose one of the Wannier functions
to have inversion symmetry about the origin, while the other three
have inversion symmetry about other Wyckoff positions (those
corresponding to the other three bond centers), and the remaining
Wyckoff positions (tetrahedral and octahedral interstitial positions)
are unoccupied.\cite{kohn73}  This would have been hard to guess based
on symmetry alone (although it is natural from a chemical point of
view).  Because each Wannier function does have its own inversion
symmetry, it turns out that $\Omega_{\rm D}$ does vanish for Si.
However, $\Omega_{\rm OD}\ne0$.  The contribution to
$\Omega_{\rm OD}$ from a given pair $\{mn\}$ of Wannier functions
is related to the matrix elements $\langle{\bf R}m\vert{\bf r}
\vert{\bf 0}n\rangle$.  These matrix elements can be shown to
vanish if, in addition to obeying inversion symmetry individually,
the two Wannier functions are translational images of one another;
but this is certainly not generally the case.  (In the language of
Appendix \ref{app:geometry}, the fact that $\Omega_{\rm OD}\ne0$
for Si is related to the fact that the Berry curvature tensor
does not vanish for this system.)

Finally, in some cases it might be possible to choose all the
Wannier functions to have definite symmetry under inversion, but the
solution that minimizes $\Omega$ may spontaneously break the
inversion symmetry.  Some cases of this sort are discussed in
Secs.\ \ref{sec:c2h4} and \ref{sec:licl} below.

\subsubsection{Molecular supercells and single k-point sampling}
\label{sec:molecsingle}

In the context of plane-wave pseudopotential and related
approaches, it is common to study molecules or clusters in an
artificial periodic superlattice arrangement.\cite{explanslab}
In such a case, a single k-point (usually ${\bf k}_0=\Gamma$) 
sampling of the Brillouin zone suffices for conventional
quantities such as energies, forces, and charge densities, since
the errors in these quantities will be exponentially small as long
as the overlap between wavefunctions in neighboring supercells is
negligible.  However, under the same conditions, the calculation
of $\Omega$ using our approach introduces small errors that
nevertheless scale only as $L^{-2}$, where $L$ is the supercell
dimension (see, e.g., Sec.\ \ref{sec:c2h4}).  The problem
essentially arises from the use of the simplest finite-difference
representation of $\nabla_{\bf k}$, involving only
nearest-neighbor k-points (see Appendix \ref{app:kmesh}).
If higher accuracy is need,
this problem can be overcome in either of two ways:
(i) by using the solution at ${\bf k}_0$ to construct solutions
on a denser mesh of k-points,
$u_{\bf k}({\bf r})=u_{{\bf k}_0}({\bf r})\,
\exp[i({\bf k}_0-{\bf k})\cdot{\bf r}]$,
being sure to take the discontinuity of
$({\bf k}_0-{\bf k})\cdot{\bf r}$ near the supercell boundary
where $u_{{\bf k}_0}({\bf r})$ is negligible;
or (ii), construct periodic functions $\widetilde x(\bf r)$,
$\widetilde y(\bf r)$, $\widetilde z(\bf r)$ such that
$\widetilde x=x$, $\widetilde y=y$, $\widetilde z=z$
in the molecular region, with (possibly smoothed) discontinuities
at the supercell boundaries, and then apply the theory of
Appendix \ref{app:realmin} to the matrices $X$, $Y$, $Z$ computed
as $X_{mn}=\langle u_{m{\bf k}_0}|\widetilde x|
u_{n{\bf k}_0}\rangle$, etc.  Approach (i) is a ``quick fix''
requiring very little reprogramming, while approach (ii) is
preferable in principle.

It is also common practice to use single k-point sampling for
supercell calculations on extended systems, provided that the
supercell is sufficiently large in all three dimensions.  In such
cases, our procedure can again be applied, but it should be kept
in mind that the convergence of $\Omega$ with supercell size
should be expected to be slower than the convergence of total
energies and forces.  Moreover, the electronic polarization that
would be computed from the sum of our Wannier centers is not
guaranteed to be exactly identical to the one that would be
computed from the Berry-phase formula\cite{ksv,parritwo}
\begin{equation}
{\bf P}_{\rm el}\cdot{\bf G}\,=\,{-2e\over V}\,{\rm Im\,ln\,det}\,
\langle\,u_{m{\bf k}_0}\,\vert\,e^{-i{\bf G}\cdot{\bf r}}\,\vert
\,u_{n{\bf k}_0}\,\rangle \;\;. 
\label{eq:parri}
\end{equation}
used in recent molecular-dynamics simulations of infrared
absorption spectra.\cite{parritwo} However, the two should be very
close, and should become identical in the limit of large supercell
size.

%----------%
\subsection{Steepest-descent minimization}
\label{sec:sd}
%----------%

\subsubsection{Algorithm}
\label{sec:sdalg}

In order to minimize the spread functional $\Omega$ by steepest
descents, we make small updates to the unitary matrices, as in
Eq.\ (\ref{eq:rot}), choosing
\begin{equation}
dW^{\bf(k)} = \epsilon \, G^{\bf(k)}
\label{eq:step}
\end{equation}
where $\epsilon$ is a positive infinitesimal.  We then have,
to first order in $\epsilon$,
\begin{eqnarray}
d\Omega &=& \sum_{\bf k} {\rm tr}\,[G^{\bf(k)}dW^{\bf(k)}] \nonumber \\
&=& - \epsilon \sum_{\bf k} \Vert G^{\bf(k)} \Vert^2 \;\;,
\label{eq:down}
\end{eqnarray}
where $\Vert A\Vert^2=\sum_{mn}|A_{mn}|^2$ and we have made use of
$G^\dagger=-G$.  Thus, use of Eq.\ (\ref{eq:step}) is guaranteed to
make $d\Omega<0$, i.e., to reduce $\Omega$.

In practice, we take a fixed finite step with $\epsilon=\alpha/4w$,
where $w=\sum_{\bf b} w_b$, so that
\begin{equation}
\Delta W^{\bf(k)} = {\alpha\over w} \sum_{\bf b} w_b \,
\left( {\cal A}[R^{\bf(k,b)}] - {\cal S}[T^{\bf(k,b)}] \right)
\;\;.
\label{eq:fstep}
\end{equation}
The wavefunctions are then updated according to the matrix
$\exp[\Delta W^{\bf(k)}]$, which is unitary because $\Delta W$ is
antihermitian.  The choice of prefactor above is designed so that
in the single-band case, and for simple k-meshes (e.g., simple
cubic), the ``highest-frequency mode'' associated with phase
rotations is just marginally stable with the choice $\alpha=1$.
That is, if one starts with the true solution and rotates the
phases of the wavefunctions on all k-points simultaneously by an
angle $\pm\gamma$, with the opposite sense of rotation on
nearest-neighbor k-points, then from Eq.\ (\ref{eq:qdef}) $\Delta
q^{\bf(k,b)}=\pm2\gamma$ on every link, and the above choice of
$\Delta W$ exactly returns the system to the solution if
$\alpha=1/2$, and is marginally unstable at $\alpha=1$.
We find that $\alpha=1$ is still a safe choice for all the 
systems studied; more efficient strategies, if needed, can also
be implemented straightforwardly, using, e.g., a conjugate-gradient
approach in composing subsequent descent directions, and by choosing 
at each step the optimal $\alpha$ for the line minimization.

It should be noted that the evolution towards the minimum requires only
the relatively inexpensive updating of the unitary matrices, and not 
of the wavefunctions, as follows.  We choose a reference set of
Bloch orbitals $\vert u_{n\bf k}^{(0)}\rangle$ and compute
once and for all the inner-product matrices
\begin{equation}
M_{mn}^{(0)\bf(k,b)} = \langle u_{m\bf k}^{(0)}
\vert u_{n,\bf k+b}^{(0)}\rangle \;\;.
\label{eq:m0def}
\end{equation}
We then represent the $\vert u_{n\bf k}\rangle$ (and thus,
indirectly, the Wannier functions) in terms of the
$\vert u_{n\bf k}^{(0)}\rangle$ and a set of unitary
matrices $U_{mn}^{\bf(k)}$,
\begin{equation}
\vert u_{n{\bf k}}\rangle =
\sum_m U_{mn}^{(\bf k)}\,\vert u_{m{\bf k}}^{(0)}\rangle \;\;.
\label{eq:uudef}
\end{equation}
We begin with all the $U_{mn}^{\bf(k)}$ initialized to $\delta_{mn}$.
Then, each step of the steepest-descent
procedure involves calculating $\Delta W$ from
Eq.\ (\ref{eq:fstep}), updating the unitary matrices according
to
\begin{equation}
U^{\bf(k)} \;\rightarrow\; U^{\bf(k)} \, \exp[\Delta W^{\bf(K)}]
\;\;,
\label{eq:uupdate}
\end{equation}
and then computing a new set of $M$ matrices according to
\begin{equation}
M^{\bf(k,b)} = U^{\bf(k)\dagger} \, M^{(0)\bf(k,b)} \, U^{\bf(k+b)}
\;\;.
\label{eq:newM}
\end{equation}
The cycle is then repeated until convergence is obtained.  Note
that the exponential in Eq.\ (\ref{eq:uupdate}) is a matrix
operation, which we perform by transforming to a diagonal
representation of $\Delta W$ and back again.

Typically, we prepare a set of reference Bloch orbitals
$\vert u_{n\bf k}^{(0)}\rangle$ by projecting from a set of
initial trial orbitals $g_n(\bf r)$ corresponding to some
rough initial guess at the Wannier functions.  For example, for
these $g_n(\bf r)$ we have used Gaussian functions centered at or
near mid-bond positions.  The initialization
procedure involves first projecting onto Bloch states of the set of
bands at wavevector $\bf k$,
\begin{equation}
\vert\phi_{n\bf k}\rangle=\sum_m\vert\psi_{m\bf k} \rangle
\langle\psi_{m\bf k}\vert g_n\rangle \;\;.
\label{eq:proj}
\end{equation}
Since these are not orthonormal, we then perform a symmetric
orthonormalization to form a set of
\begin{equation}
\vert\widetilde{\phi}_{n\bf k}\rangle = \sum_m (S^{-1/2})_{mn}
\vert\phi_{m\bf k}\rangle
\label{eq:symm}
\end{equation}
(where $S_{mn}=\langle\phi_{m\bf k}\vert\phi_{n\bf k}\rangle$),
and finally convert to cell-periodic functions via
\begin{equation}
u_{n\bf k}^{(0)}({\bf r})=e^{-i\bf k\cdot r}
\widetilde{\phi}_{n\bf k}({\bf r}) \;\;.
\label{eq:phitou}
\end{equation}
(In practice, the above steps are combined.) This procedure is
similar in principle to the one introduced by Satpathy and
Pawlowska,\cite{satpathy} although it differs in that we do the
orthonormalization in k-space.  We then use this set of reference
Bloch orbitals as a starting point for the steepest-descent
procedure.  In practice, we find that this starting guess is
usually quite good, as will be shown for the cases of Si and GaAs 
in Sec.\ \ref{sec:results}.

\subsubsection{False local minima}
\label{sec:falsemin}

We have also carried out tests in which we initialize the
steepest-descent procedure with more arbitrary starting guesses.
For example, we have let the starting $u_{n\bf k}^{(0)}$ consist of
energy-ordered Hamiltonian eigenstates with quasi-random phases, as
in the typical output of a band-structure code.  We have also
tried applying a completely random phase rotation to 
each $u_{n\bf k}^{(0)}$
individually, or a random $J\times J$ unitary rotation to the set
of $u_{n\bf k}^{(0)}$ at each k.  With such starting guesses, we find
that while the steepest-descent procedure sometimes does lead to the
desired global minimum, it often can get stuck in local minima.
That is, we find that the spread functional $\Omega$, viewed as a
function of the set of $U_{mn}^{\bf(k)}$, does typically have false
local minima that must be avoided.

We find that this problem is {\it not} associated with
the presence of a large number of bands, and in particular it
never arose in the case of $\Gamma$-only sampling in orthorhombic
supercells (say, even for a disordered 64-atom cell of Si). 
Instead, it tends to be associated with finer
k-point meshes, regardless of whether one is treating single or
multiple bands.  In the case of multiple bands, some
Wannier functions at the local minimum have an almost normal 
behavior, with reasonable spread, while other functions are abnormal, 
with spreads an order of magnitude larger than expected.  
Moreover, we found all the false minima to be characterized by
Wannier functions that are genuinely complex (for the true global
minimum we always found the Wannier functions to be real, apart
from a trivial overall phase).  The Wannier functions associated
with false local minima usually display erratic and unphysical
oscillations.

The problem appears to lie in the possibility
of making inconsistent choices in the branch cuts when evaluating the
logarithms of complex argument in (\ref{eq:qdef}).  In a naive
implementation, the branch cuts are simply chosen so that
$|q_n^{\bf(k,b)}| \le \pi$.  At a good global minimum, all of the
$|q_n^{\bf(k,b)}| \ll \pi$, while at a false local minimum some of the
$|q_n^{\bf(k,b)}|$ approach $\pi$.  We find that the system can
be excited out of these local minima by an iterative process of
switching the branch cut by 2$\pi$ for the few $q_n$'s with the
largest magnitudes, followed by further steepest-descent minimization,
and that such a process almost always leads fairly quickly into the
basin of attraction of the global minimum.

Moreover, we have never observed the system to become trapped in a
false local minimum when starting from reasonable trial projection
functions, Eqs.\ (\ref{eq:proj}-\ref{eq:phitou}).  If physically
motivated trial functions are not available, we find that another
effective heuristic approach is to use as trial projection
functions Gaussians that are either centered at random locations,
or on arbitrary meshes in the unit cell.

In summary, while false local minima can occur in our minimization
scheme, they do not seem to pose a problem in practice.

% -------------------------------------------------------------------

\section{Properties of optimally-localized Wannier functions}
\label{sec:miscprop}

% -------------------------------------------------------------------

%----------%
\subsection{Asymptotic localization properties}
\label{sec:localization}
%----------%

Following from the early work of Kohn,\cite{kohn59} it is generally
expected that Wannier functions can be chosen to have exponential
localization.  While it is not the purpose of the present
work to study questions of exponential decay in the tails of the
Wannier functions, we nevertheless give a brief discussion of
these issues here.

Kohn \cite{kohn59} proved the existence of exponentially localized
Wannier functions for the case of an isolated band in 1D, for a
crystal with inversion symmetry.  However, the method does not
easily generalize.  Blount demonstrated the analyticity of the
Bloch functions for the single-band case in 3D,\cite{blount}
and claimed (end of Sec.\ 5 of Ref.\ \onlinecite{blount}) that
this would imply the exponential localization of the Wannier
functions (see also Ref.\ \onlinecite{weinreich});
but this claim was later shown to be faulty by Nenciu
(footnote on first page of Ref.\ \onlinecite{nenciu83}),
who pointed out the global topological aspects of the problem.
Des Cloizeaux proved the exponential localization of the
band projection operator $P$ of Eq.\ (\ref{eq:P}) for an arbitrary
set of composite bands in 3D.\cite{cloizeauxa}  Unfortunately,
this does not immediately imply that the Wannier functions are
exponentially localized (although the converse would follow).
In a following paper, des Cloizeaux was able to prove the possibility
of choosing exponentially localized Wannier functions for
an isolated band (i) in 1D generally, or (ii) in the centrosymmetric
3D case.\cite{cloizeauxb}  The summary (Sec.\ V) of
Ref.\ \onlinecite{cloizeauxb} gives a good discussion of the
difficulties and partial progress towards a solution of the general
composite-band problem.  More recently, Nenciu completed a proof
for the case of an isolated band in 3D without
centrosymmetry.\cite{nenciu83}  To our knowledge, however,
the problem remains unsolved for the general case of composite bands
in 3D.  Finally, note that some discussion of the exponential
localization of the ``generalized Wannier functions'' defined for
the cases of surfaces and defects has been given in
Refs.\ \onlinecite{onffroy,rehr,geller,nenciu93}.

It is natural to speculate that the ``optimally localized'' Wannier
functions that are obtained by minimizing the spread functional of
Eq.\ (\ref{eq:om1}) are exponentially localized.  Actually, one
should distinguish between a ``weak conjecture'' that the optimally
localized Wannier functions have exponential decay, and a ``strong
conjecture'' that they have the same exponential decay as that of the
band projection operator $P$.  At the present time, we can only
speculate that in 3D, the weak conjecture, at least, will hold.

In 1D, we are on firmer footing.  As shown in Secs.\ \ref{sec:rspace}
and \ref{sec:1D},
the functions that are obtained by minimizing Eq.\ (\ref{eq:om1})
correspond, in principle, with those considered by previous authors,
and for which exponential localization has been
demonstrated.\cite{kohn59,blount,kivelson,niu}  In particular,
we have shown in Sec.\ \ref{sec:rspace}
that these will be eigenfunctions of the band-projected
position operator $PxP$; Niu has given a simple and elegant argument,
based on this fact alone, from which one may conclude that
the Wannier functions decay faster than any power.\cite{niu} From
this point of view, the essential difficulty in 3D is that the Wannier
functions can no longer generally be chosen to be eigenfunctions
of all three band-projected position operators simultaneously.

Returning to the general 3D case, we find that it is not easy to
carry out numerical tests of exponential localization using the
present method, which is based on discretization in k space.  The
Wannier functions that we obtain are thus not truly localized,
being instead artificially periodic with a periodicity inversely
proportional to the mesh spacing.

% -------------------------------------------------------------------

\subsection{Conjecture: optimally localized Wannier functions are real}
\label{sec:real}

% -------------------------------------------------------------------

It seems not to be widely appreciated that the Wannier functions
$w_n(\bf r)$ can always be chosen real.  This depends only on
the Hamiltonian $H=p^2/2m+V({\bf r})$ being hermitian, and not
on any symmetry of the (real) potential $V(\bf r)$.  Indeed, from
Eq.\ (\ref{eq:wdef}) it is clear that one only needs to choose
\begin{equation}
u_{n{\bf k}}({\bf r})=u_{n,{\bf-k}}^*(\bf r)
\label{eq:ureal}
\end{equation}
to insure that the Wannier functions $w_n(\bf r)$ are real.
This condition is automatically satisfied if one starts with initial
Wannier functions projected from real trial functions, as discussed
at the end of the previous subsection; alternatively, it can be
imposed by hand.  From Eq.\ (\ref{eq:mdef}), condition
(\ref{eq:ureal}) implies that $M_{mn}^{\bf(k,b)} =
M_{mn}^{\bf(-k,-b)*}$, which in turn implies $G_{mn}^{\bf(k)} =
G_{mn}^{\bf(-k)*}$, so that Eq.\ (\ref{eq:ureal}) continues to be
satisfied during the steepest-descent update procedure.  In this
way, one will eventually arrive at a set of maximally localized
real Wannier functions.  (Similarly, working in real space,
it is easy to see from Appendix \ref{app:realmin} that a real initial
guess will result in a set of real optimally-localized solutions.)

We conjecture that a stronger result is true: namely, that the
optimally localized Wannier functions are always real (apart
from a trivial overall phase of each Wannier function).  That
is, we conjecture that the minimum that is arrived at subject to
the constraint (\ref{eq:ureal}) is actually the global minimum.

We have not found a proof of this conjecture, but it is supported
by our empirical experience.  More precisely, in the tests
to be reported in Sec.\ \ref{sec:results}, we find that whenever
we arrive at the global minimum, the Wannier functions
always turn out to be real, apart from a trivial overall phase.
(However, we do find that the Wannier functions are typically
complex at false local minima, as discussed in
Sec.\ \ref{sec:falsemin}.)

% -------------------------------------------------------------------

\section{Results}
\label{sec:results}

\subsection{S\lowercase{i}}
\label{sec:si}

For Si, the four occupied valence bands have to be taken together
as a single composite group, because of degeneracies between the
bottom two bands at X, and between the top three bands at
$\Gamma$.  Thus, we take $J=4$ and look for a set of four Wannier
functions per primitive unit cell.  These are expected to be
centered on the bond centers, and to have roughly the character of
$\sigma$-bond orbitals, i.e., even linear combinations of the two
$sp^3$ hybrids projecting toward the bond center from the two
neighboring atoms.\cite{kohn73} Wannier functions of this type have
been computed previously by a variety of
methods.\cite{satpathy,kane,tejedor,pederson,fernandez}  It is
tempting to imagine that the requirement of spanning the given set
of valence bands, together with the symmetry requirement that each
Wannier function has the expected inversion, mirror, and three-fold
rotational symmetries about its corresponding bond center, might be
enough to uniquely determine the Wannier functions.  We emphasize
that this is not the case, and we proceed to determine the
particular set of Wannier functions that minimize the spread
functional $\Omega$.

Our calculations are carried out within the local-density approximation
to Kohn-Sham density-functional theory,\cite{dft-lda}
using a standard plane-wave
pseudopotential approach and an all-bands conjugate-gradient
minimization.\cite{payne}  We have used norm-conserving 
pseudopotentials\cite{lin} in the Kleinman-Bylander representation, 
with plane-wave cutoffs ranging from 200 eV to 650 eV, depending
on the systems studied. The sampling of the Brillouin-zone
is performed with equispaced Monkhorst-Pack grids\cite{monkpack}
that have been offset in order to include $\Gamma$.
Since the crystal is fcc in real space, the grid is bcc in k-space, and
we use the simplest possible finite-difference representation of
$\nabla_{\bf k}$ using only the $Z$=8 nearest neighbors of each k-point
(see Appendix \ref{app:kmesh}).
The computed Bloch functions are stored to disk, and
the construction of the Wannier functions is carried out as a separate,
post-processing operation.

Table \ref{tab:si-kconverg} shows the convergence of the spread
functional and its various contributions as a function of the
density of the k-point mesh used. We confirm that $\Omega_{\rm D}$
does vanish (to machine precision) as expected from the presence of
inversion symmetry, as discussed in Sec.\ \ref{sec:inversion}.
Since $\Omega_{\rm I}$ is invariant, the minimization of $\Omega$
reduces to the minimization of $\Omega_{\rm OD}$.  For each k-point
set, the minimization was initialized by starting with trial
Gaussians of width (standard deviation) 1 \AA\ located at the bond
centers.  We find that for the case of crystalline Si, these
provide an excellent starting guess; for the $8\times 8\times 8$
case, for example, we find an initial $\Omega_{\rm D}$=0 and
 $\Omega_{\rm OD}$=0.565, whereas at the minimum $\Omega_{\rm OD}$
is 0.520. Had we started with the random phases provided by the
ab-initio code, we would have obtained an initial $\Omega_{\rm
D}$=622.1 and $\Omega_{\rm OD}$=42.3.  We find that typically 20
iterations are needed to converge to the minimum with good
accuracy, starting with the initial choice of phases given by the
Gaussians, and using a simple fixed-step steepest-descent
procedure.  Starting with a set of randomized phases requires
roughly one order of magnitude more iterations, and adds the
possibility that the system may get trapped for a while in some
local minimum.  As previously pointed out, the evolution does not
require additional scalar products between Bloch orbitals, and so
it is in any case pretty fast.  Because of symmetry, the Wannier
centers do not move during the minimization procedure, and the
spreads of the four Wannier functions remain identical with each
other.

\begin{table}
\caption{Minimized localization functional $\Omega$ in Si, and its
decomposition into invariant, off-diagonal, and diagonal parts,
for different k-point meshes (see text).  Units are \AA$^2$.}
\begin{tabular}{ldddd}
k set & $\Omega$ & $\Omega_{\rm I}$ & $\Omega_{\rm OD}$ &
   $\Omega_{\rm D}$ \\
\tableline
$1\times 1\times 1$ & 2.024 & 1.999 & 0.025 & 0 \\
$2\times 2\times 2$ & 4.108 & 3.707 & 0.401 & 0 \\
$4\times 4\times 4$ & 6.447 & 5.870 & 0.577 & 0 \\
$6\times 6\times 6$ & 7.611 & 7.048 & 0.563 & 0 \\
$8\times 8\times 8$ & 8.192 & 7.671 & 0.520 & 0 \\
\end{tabular}
\label{tab:si-kconverg}
\end{table}

What is perhaps most striking about Table \ref{tab:si-kconverg} is
that $\Omega_{\rm I} \gg \Omega_{\rm OD}$; and while $\Omega$
converges fairly slowly with k-point density, this poor convergence
is almost entirely due to the $\Omega_{\rm I}$ contribution.
Incidentally, since the $\Omega_{\rm I}$ contribution is
gauge-invariant, it can be calculated once and for all at the
starting configuration, for any given k-point set; the quantities
that are actually minimized are $\Omega_{\rm D}$ and $\Omega_{\rm
OD}$.  The former vanishes at the minimum, and the latter is found
to converge quite rapidly with k-point sampling.  It would be
interesting to explore whether use of a higher-order
finite-difference representation of $\nabla_{\bf k}$ might improve
this convergence, especially that of $\Omega_{\rm I}$, but we have
not investigated this possibility.

\begin{figure}
\epsfxsize=3.2 truein
\centerline{\epsfbox{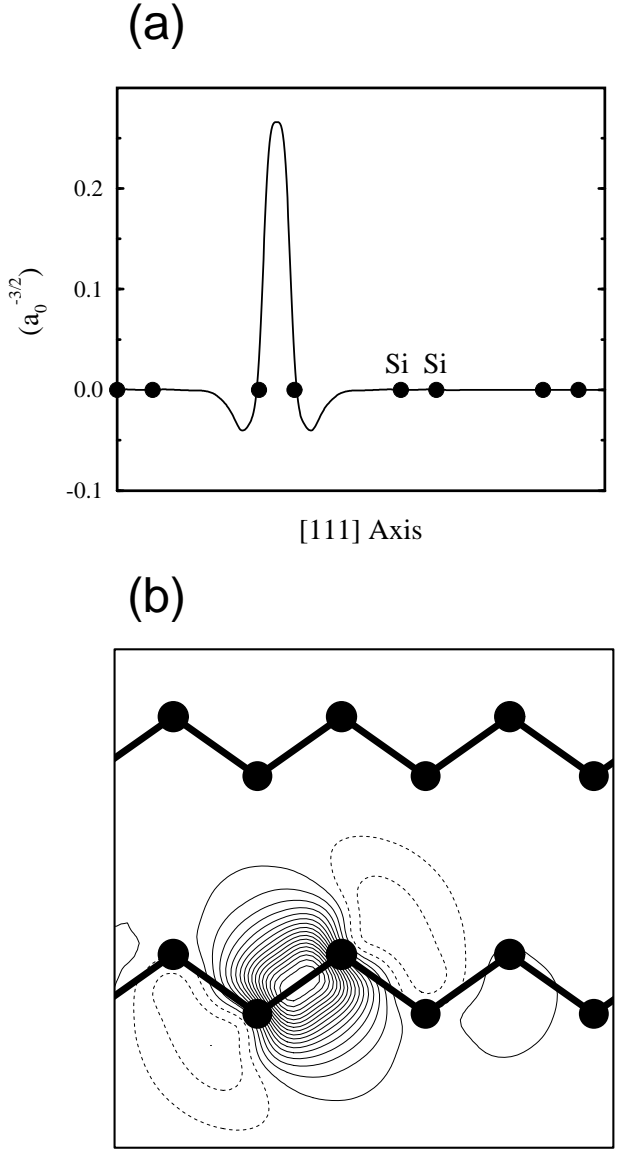}}
\vskip 0.2truein
\caption{Maximally-localized Wannier function in Si, for the 
$8\times 8\times 8$ k-point sampling. (a) Profile along the Si-Si
bond. (b) Contour plot in the (110) plane of the bond chains.
The other Wannier functions lie on the other three tetrahedral bonds
and are related by tetrahedral symmetries to the one shown.}
\label{fig:si-cplot}
\end{figure}

In Fig.\ \ref{fig:si-cplot}, we present plots showing one of these
maximally-localized Wannier functions in Si, for the $8\times
8\times 8$ k-point sampling.  The other three are identical
(related to the first by the tetrahedral symmetry operations) and
are located on the other three tetrahedral bonds.  Each displays
inversion symmetry about its own bond center, and it is real apart
from an overall complex phase.  Again, all these properties are not
trivial, and would not be satisfied by a generic choice of phases.
(Our initial guess based on Gaussians centered in the middle of the
bonds does insure all these properties, but without optimizing the
localization.)

From an inspection of the contour plot it becomes readily apparent
that
the Wannier functions are essentially confined to the first unit
cell, with very small (and decreasing) components in
further-neighbor shells.  The general shape corresponds to a
chemically intuitive view of $sp^3$ hybrids overlapping along the
Si-Si bond to form a $\sigma$ bond-orbital, with the smaller lobes
of negative amplitude clearly visible in the back-bond regions.
These results clearly illustrate how the Wannier functions can
provide useful intuitive understanding about the formation of
chemical bonds.

\subsection{G\lowercase{a}A\lowercase{s}}
\label{sec:gaas}

In GaAs the lower valence band is never degenerate with the other
(top) three valence bands, and thus several possibilities arise:
(a) We can treat the four bands as a group, as was done for silicon,
obtaining solutions that are very similar to the Si case, except for
the loss of inversion symmetry about the bond centers.
(b) We can deal separately with the bottom band and the top three
bands; the latter would be considered as a group, while the former
is a single isolated band. The solution at the minimum should
resemble atomic orbitals for the more electronegative species (the
As anion), in the form of three $p$ orbitals and one $s$ orbital
respectively.  (c) Finally, it might be interesting to consider the
case in which the four bands are treated together, but using the
solution of the $\Omega$ minimization for the 1-band and 3-band
cases, without proceeding further with the minimization.  This does
not correspond to a true minimum for the 4-band $\Omega$ surface,
but just to a stationary (saddle) point.

\begin{table}
\caption{Minimized localization functional $\Omega$ in GaAs, and its
decomposition into invariant, off-diagonal, and diagonal parts,
for different k-point meshes, together with the relative position
$\beta$ of the centers along the Ga-As bond (see text).  Units for the
$\Omega$'s are \AA$^2$.}
\begin{tabular}{lddddd}
k set & $\Omega$ & $\Omega_{\rm I}$ & $\Omega_{\rm OD}$ &
    $\Omega_{\rm D}$ & 
$\beta$ \\
\tableline
$1\times 1\times 1$ & 2.217 & 2.088 & 0.125 & 0.0035 & 0.593 \\
$2\times 2\times 2$ & 4.409 & 3.898 & 0.503 & 0.0078 & 0.602 \\
$4\times 4\times 4$ & 6.785 & 6.170 & 0.610 & 0.0055 & 0.613 \\
$6\times 6\times 6$ & 7.982 & 7.386 & 0.590 & 0.0058 & 0.616 \\
$8\times 8\times 8$ & 8.599 & 8.038 & 0.555 & 0.0059 & 0.617 \\
$12\times 12\times 12$ & 9.146 & 8.635 & 0.504 & 0.0061 & 0.617 \\
\end{tabular}
\label{tab:gaas-kconverg}
\end{table}

Starting with the case in which all the four bands are treated as a
group, we show in Table \ref{tab:gaas-kconverg} the convergence of
the spread functional and its various contributions as a function
of the density of the k-point sampling. In analogy with the case of
Si, the procedure is initialized using trial Gaussians of width 1
\AA, centered in the middle of the bonds; this is again a very good
starting guess, and (for the $8\times 8\times 8$ mesh), gives an
initial $\Omega_{\rm D}$=0.1164 and $\Omega_{\rm OD}$=0.593, that
are reduced to 0.0059 and 0.555 respectively by the minimization
procedure.  As it was the case for Si, k-point convergence is
fairly slow, even though most of it is due to the slow convergence
of the invariant part.  On the other hand, the general shape of the
Wannier functions at the minimum is already given rather accurately
with coarser samplings (although the tails are then not so easy to
characterize, since in practice the Wannier functions are
periodically repeated in a supercell conjugate to the k-point
mesh). In particular, the k-point convergence of the Wannier
centers is quite rapid, as is evident from the last column of Table
\ref{tab:gaas-kconverg}, where we show the relative position of the
centers along the Ga-As bonds. Here $\beta$ is the distance between
the Ga atom and the Wannier center, given as a fraction of the bond
length (in Si the centers were fixed by symmetry to be in the
middle of the bond, $\beta=0.5$, irrespective of the sampling).

\begin{figure}
\epsfxsize=3.2 truein
\centerline{\epsfbox{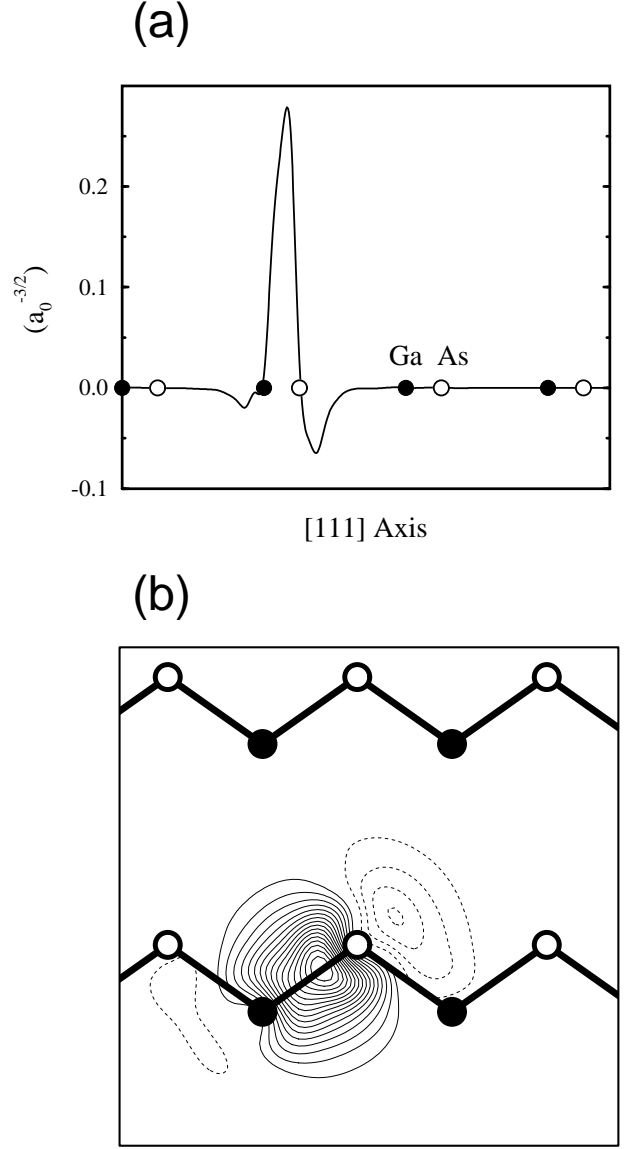}}
\vskip 0.2truein
\caption{Maximally-localized Wannier function in GaAs, for the 
$8\times 8\times 8$ k-point sampling. (a) Profile along the
Ga-As bond.  (b) Contour plot in the (110) plane of the bond chains.
The other Wannier functions lie on the other three tetrahedral bonds
and are related by tetrahedral symmetries to the one shown.}
\label{fig:gaas-cplot}
\end{figure}

In Fig.\ \ref{fig:gaas-cplot}, we present plots showing one of
these maximally-localized Wannier functions in GaAs, for the
$8\times 8\times 8$ k-point sampling. Again, at the minimum
$\Omega$, all four Wannier functions have become identical (under
the symmetry operations of the tetrahedral group), and they are
real, except for an overall complex phase. The shape of the Wannier
functions is again that of $sp^3$ hybrids combining to form
$\sigma$-bond orbitals; inversion symmetry is now lost, but the
overall shape is otherwise closely similar to what was found in
Si.  The Wannier centers are still found along the bonds, but they
have moved towards the As, at a position that is 0.617 times the
Ga-As bond distance.  It should be noted that these Wannier
functions are also very similar to the localized orbitals that are
found in linear-scaling approaches,\cite{fernandez} where
orthonormality, although not imposed, becomes exactly enforced in
the limit of an increasingly large localization region.  This
example highlights the connections between the two approaches.  The
characterization of the maximally-localized Wannier functions
indicates the typical localization of the orbitals that can be
expected in the linear-scaling approach.  Moreover, such
information ought to be extremely valuable in constructing an
intelligent initial guess at the solution of the electronic
structure problem in the case of complex or disordered systems.

\begin{table}
\caption{Localization functional $\Omega$ and its decomposition in
invariant, off-diagonal, and diagonal parts, for the case of GaAs
(units are \AA$^2$).  The bottom valence band, the top three
valence bands, and all four bands are separately included in the
minimization. The star ($^{\star}$) refers to the case in which the
minimization is not actually performed, and the solution for the
1-band and 3-band cases is used.  Sampling is performed with a
$8\times 8\times 8$ mesh of k-points.}
\begin{tabular}{ldddd}
k set & $\Omega$ & $\Omega_{\rm I}$ & $\Omega_{\rm OD}$
    & $\Omega_{\rm D}$ \\
\tableline
1 band & 1.968 & 1.944 & 0 & 0.0238 \\
3 bands & 10.428 & 9.844 & 0.560 & 0.0245 \\
4 bands$^{\star}$ & 12.396 & 8.038 & 4.309 & 0.0483 \\
4 bands & 8.599 & 8.038 & 0.555 & 0.0059 \\
\end{tabular}
\label{tab:gaas-134bands}
\end{table}

As pointed out before, in GaAs we can have different choices for
the Hilbert spaces that can be considered, so we also studied the
case in which only the bottom band, or the top three, are used as
an input for the the minimization procedure.  Table
\ref{tab:gaas-134bands} shows the spread functional and its various
contributions for these different choices, where the bottom band is
first treated as isolated; next the three $p$ bands are treated as
a separate group; then these two solutions are used to construct a
four-band solution, without further minimization; and finally, this
is compared with the full four-band minimization.  In composing the
results for the 1-band and 3-band cases, we take the $1\times 1$
and $3\times 3$ unitary matrices that would give the minimum
solution for the 1- and 3-band cases, and build from them a set of
$4\times 4$ block-diagonal unitary matrices.  The 4-band $\Omega$
that is obtained is exactly the sum of the two initial $\Omega$'s.
Nevertheless, the bookkeeping changes: $\Omega_{\rm I}$ is reduced,
with an equal and opposite contribution reappearing in $\Omega_{\rm
OD}$.  (The $\Omega_{\rm D}$'s sum up exactly, as they must.) If we
then minimize this (saddle-point) solution, we recover the 4-band
minimum: the invariant part (obviously) does not change, while
$\Omega_{\rm D}$ increases to permit a larger reduction in
$\Omega_{\rm OD}$, in correspondence to an increased interband
mixing.

\begin{figure}
\epsfxsize=3.2 truein
\centerline{\epsfbox{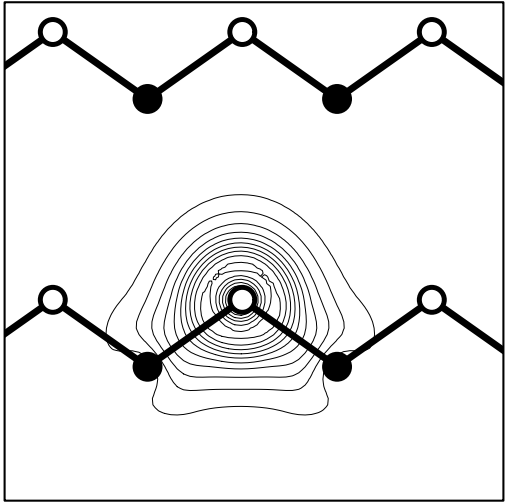}}
\vskip 0.2truein
\caption{Contour plot, in (110) plane, of the maximally-localized
Wannier function in GaAs for the $8\times 8\times 8$ k-point sampling
when only the bottom valence band is considered.}
\label{fig:gaas-cplot1}
\end{figure}

In Fig.\ \ref{fig:gaas-cplot1}, we show the contour plot for the
maximally-localized one-band Wannier function in GaAs, for the
$8\times 8\times 8$ k-point sampling. The function is again real,
and it shows the typical characteristics of an $s$ orbital centered
around the anion; the tetrahedral symmetry of the lattice deforms
the spherical orbital, introducing contributions that point along
the two bond-chains (one in the (110) plane plotted, and one
perpendicular to that plane). In the three-band case, on the other
hand, the Wannier functions resemble three orthogonal atomic $p$
orbitals.  It should be stressed that only when all the four bands
are treated simultaneously do we achieve the overall maximum
localization.  This reinforces the picture in which the maximally
localized orbitals correspond to the most natural ``chemical
bonds'' in the system.

\subsection{Molecular C$_2$H$_4$}
\label{sec:c2h4}

We have also studied the case of the ethylene molecule
(C$_2$H$_4$), in order to make the connection with some standard
chemistry concepts, and to highlight the relation of our formalism
(derived from a k-space representation of extended Bloch orbitals)
to the case of an isolated system as discussed in
Sec.\ \ref{sec:molecsingle}.  First of all, the molecule is modeled
in periodic boundary conditions, in a supercell that is large
enough to make the interaction with the periodic images negligible.
Consequently, the band dispersion becomes also negligible, and
$\Gamma$ sampling is all that is needed for total energies, forces,
and densities.  However, the spread functional is expected to
converge slightly slower with k-point sampling, as discussed in
Sec.\ \ref{sec:molecsingle}.  We thus tested several k-point
meshes.  For the single k-point case, the mesh in reciprocal space
is that formed by the $\Gamma$ point and all its periodical images,
i.e., the reciprocal lattice vectors; our formalism remains equally
applicable to such a case.  One should bear in mind that if the
supercell is not cubic, appropriate weight factors have to be added
in the calculation of the derivatives (see Appendix
\ref{app:kmesh}).

\begin{table}
\caption{Coordinates (in \AA) of the atoms and of the 6 Wannier
centers in the ethylene molecule.}
\begin{tabular}{ldddd}
 Species & x & y & z &
  \\ \tableline
H & -1.235 & 0.936 & 0.000 & \\
H &  1.235 & -0.936 & 0.000 & \\ 
H &  1.235 & 0.936 & 0.000 & \\ 
H & -1.235 & -0.936 & 0.000 & \\ 
C & 0.660 & 0.000 & 0.000 & \\ 
C & -0.660 & 0.000 & 0.000 & \\ \tableline
 WF & $\overline{r}_x$ & $\overline{r}_y$ & $\overline{r}_z$ &
  \\ \tableline
1 & -1.049 & 0.622 & 0.000 & \\
2 &  1.049 & -0.622 & 0.000 & \\ 
3 &  1.049 & 0.622 & 0.000 & \\ 
4 & -1.049 & -0.622 & 0.000 & \\ 
5 & 0.000 & 0.000 & 0.327 & \\ 
6 & 0.000 & 0.000 & -0.327 & \\ 
\end{tabular}
\label{tab:ethy1}
\end{table}

\begin{figure}
\epsfxsize=2.7 truein
\centerline{\epsfbox{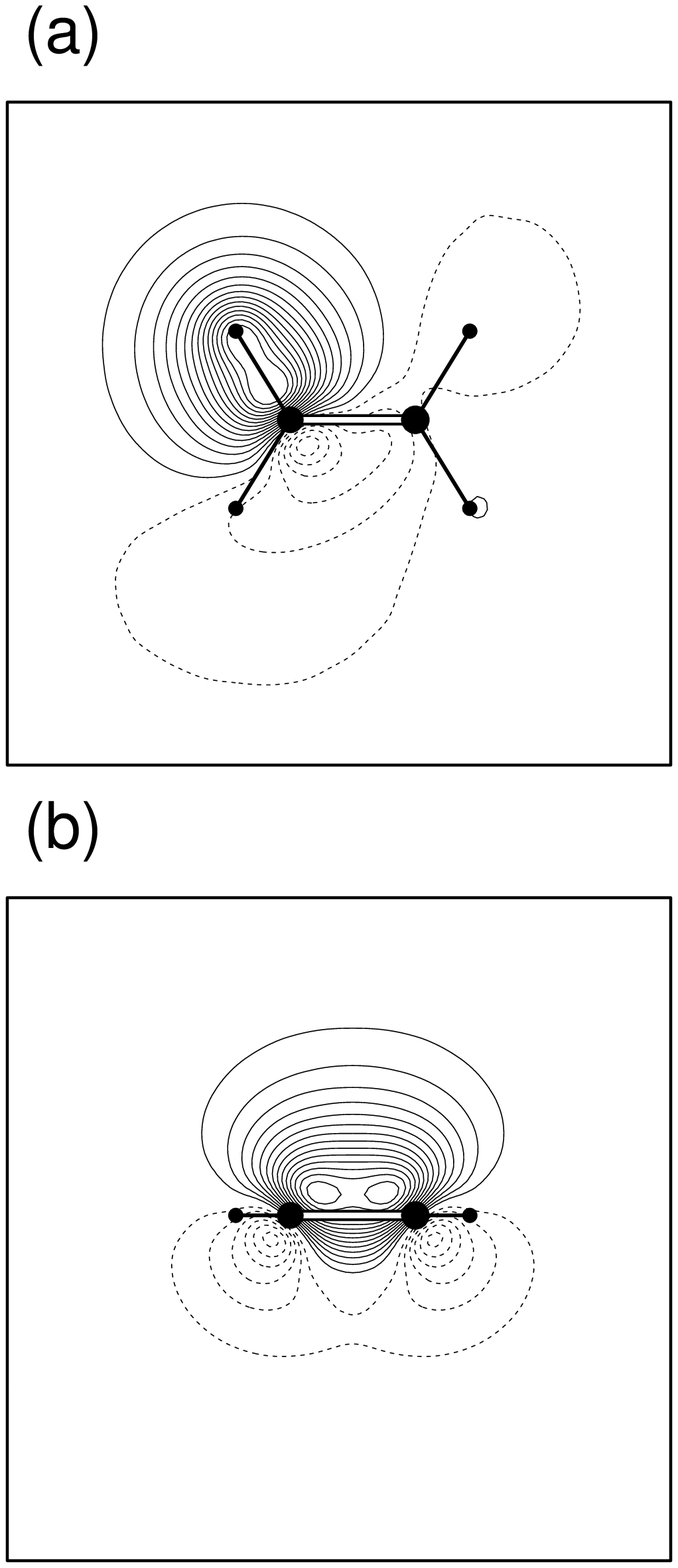}}
\vskip 0.2truein
\caption{Contour plots for the maximally-localized Wannier functions in 
ethylene, C$_2$H$_4$.
(a) One of the four C--H Wannier functions, shown in the $x-y$ plane.
(b) One of the two C=C Wannier functions, shown in $x-z$ plane.}
\label{fig:ethylene-cplot}
\end{figure}

We show in Table \ref{tab:ethy1} the coordinates for the C and H atoms
at the structural minimum, together with the Wannier centers.
We have used the local-density approximation for the
exchange-correlation functional.\cite{dft-lda} In this
molecule, there are six occupied valence eigenstates, the lowest
five being of C--H or C--C $\sigma$-bonding character, and the top
(frontier) orbital being of C--C $\pi$-bonding character.  If we treat
the lowest five bonds as a composite group, we find as expected that
the minimization of $\Omega$ leads to $\sigma$-bond-orbitals located
on each of the C--H or C--C bonds.  However, treating all six bands
together, we find that the C--C $\pi$-bonding orbital mixes strongly
with the C--C $\sigma$-bonding orbital to give two Wannier functions
that are symmetrically disposed above and below the $x-y$ plane.
Contour plots for the resulting C--H and C=C Wannier functions 
are shown in Fig. \ref{fig:ethylene-cplot}, and the locations of the
Wannier centers are reported in Table \ref{tab:ethy1}.  The picture
that emerges from this ``natural'' symmetry breaking of the planar
geometry is just the Lewis picture of the C=C double bond.

\begin{table}
\caption{The functional $\Omega$ and its decomposition, with increasing
k-point sampling, for ethylene (units are \AA$^2$).}
\begin{tabular}{ldddd}
k set & $\Omega$ & $\Omega_{\rm I}$ & $\Omega_{\rm OD}$ &
    $\Omega_{\rm D}$ \\
\tableline
$1\times 1\times 1$ & 4.041 & 3.657 & 0.384 & 0 \\ 
$2\times 2\times 2$ & 4.503 & 4.124 & 0.380 & 6$\times$ 10$^{-7}$ \\ 
$3\times 3\times 3$ & 4.600 & 4.222 & 0.377 & 3$\times$ 10$^{-7}$ \\ 
\end{tabular}
\label{tab:ethy2}
\end{table}

In our calculations we have used a cubic supercell of side 7 $\AA$;
this gives to each band a dispersion that is always smaller than
0.02 eV, and that originates from the interaction with the
superperiodic images.  Increasing the k-point sampling has
negligible effects on the equilibrium positions of the C and H
atoms and on the location of the Wannier centers.  But it does
still affect the localization functional, which displays a
slower convergence with respect to the number of k-points used
(although much faster than was the case for Si or GaAs).  The
results are summarized in Table \ref{tab:ethy2}, where we show the
$\Omega$ contributions for the maximally-localized Wannier
functions with increasing k-point sampling.  It is readily seen
that the slow convergence is coming mostly from the invariant part
of the functional; a finer k-point mesh provides both a more
detailed sampling of the Brillouin Zone and a more accurate
calculation of the gradients.

\subsection{L\lowercase{i}C\lowercase{l}}
\label{sec:licl}

It is also interesting to look at a more ionic system, to understand 
the effect of electronegativity and band gap on the location
and localization of the Wannier functions. We have studied rocksalt
LiCl, treating all four valence bands (roughly Cl 3$s$ and 3$p$)
as a unit, and again using an $8\times 8\times 8$ k-point sampling.

One could expect the Wannier functions to localize much more strongly
around the anion than was the case for GaAs, and indeed this is
what we find.  However, we also find that the Wannier functions can
reduce $\Omega$ further by mixing to form $sp^3$ hybrids, sitting on
the vertices of a tetrahedron centered around the Cl atom, with
each center at a distance of 0.449 \AA\ from the Cl (the Li-Cl
distance being 2.57 \AA).
We anticipated that these hybrids might prefer to align along the
$\{111,\bar1\bar11,\bar11\bar1,1\bar1\bar1\}$ or
$\{11\bar1,1\bar11,\bar111,\bar1\bar1\bar1\}$ sets of directions;
if this were the case, the choice between the two sets (two degenerate
global minima of $\Omega$) would constitute a kind of unphysical or
``anomalous'' symmetry breaking from cubic to tetrahedral.
Instead, we find that $\Omega$ is, at least to our machine precision,
rotationally invariant with respect to the orientation of the $sp^3$
hybrids, just as would be the case for an isolated Cl$^-$ ion in
free space.  This implies that the tetrahedron of the Wannier
centers around each Cl atom is free to rotate without any discernible
decrease of localization.

Finally, consistent with the idea that a larger gap is linked to a
higher degree of localization, we find a total $\Omega$=4.159 \AA$^2$,
with $\Omega_{\rm I}$=3.354, $\Omega_{\rm OD}$=0.805 and
$\Omega_{\rm D}$=1.2$\times$10$^{-5}$ \AA$^2$.

% -------------------------------------------------------------------

\section{Discussion}
\label{sec:discuss}

We have discussed a technique for obtaining a set of well-localized
Wannier functions for a given band or composite set of bands in
a crystalline solid.  We have in mind several kinds of applications
for this method.

First, we believe that this approach may help to obtain chemical
intuition about the nature of chemical bonds in solids, and to
characterize trends in bonding properties within classes of
solids.  As emphasized in the introduction, the Wannier functions
defined here are the natural generalization of the concept of
``localized molecular orbitals''
\cite{boys,foster,edmiston,boysr2,weinstein,leonard} to the case
of solids.  As illustrated in the examples of GaAs and ethylene
(C$_2$H$_4$) above, the determination of the Wannier functions can
give chemical intuition into the nature of the bond orbitals of the
material, including the spontaneous symmetry breaking that occurs
in the Lewis picture of a double or triple bond.  We also suspect
that it may be instructive to generate, characterize and plot the
Wannier functions across a series of compounds, e.g., for II-VI
semiconductors as one varies from wide- to narrow-gap members, or
in cubic perovskites of varying composition.  Moreover, as
emphasized by Hierse and Stechel,\cite{stechel} the Wannier functions 
may be transferable to a considerable degree for similar bonds in
different chemical systems (for example, for C-H or C-C bonds in a
variety of hydrocarbons).  It should be noted, however, that this
is even more likely to be true for non-orthogonal Wannier-like
functions,\cite{stechel} as opposed to the orthogonal ones studied
here.

Second, it is possible that the Wannier functions may prove
suitable as a basis for use in constructing theories of interacting
or strongly-correlated electron systems.  For example, it might
be possible to build good approximate correlated wavefunctions from
sums of Slater determinants of the Wannier functions. For this
purpose, one would clearly need to choose a set of bands that
includes some low-lying unoccupied states of the one-particle
mean-field Hamiltonian.  Similarly, it might be possible to
build accurate model Hamiltonians for for magnetic systems, or for
transport properties of metals.  (Again, for metals it would appear
necessary to choose a composite group of bands that brackets the
Fermi level, and to specify the occupation as a kind of density
matrix in the Wannier indices.)

Third, the present scheme might prove useful for predicting the
suitability of linear-scaling methods for different kinds of
insulating materials.  Since the linear-scaling methods
\cite{gallireview} depend strongly on the localization properties
of the Wannier functions (or, closely related, the density matrix),
the present scheme might be a simple and useful way to characterize
the degree of localization for a given target material.
This information might then help predict whether the material is
a good candidate for a linear-scaling method; and if so, what type
of linear-scaling method is likely to work best, and what
real-space cutoff parameter is likely to be required.

Finally, an important feature of the present approach is
that it generates a list of the locations of the Wannier centers.
This information alone can often be of crucial importance.
In fact, we envisage a number of interesting applications in which
one essentially throws away all other information about the Wannier
functions, keeping only their locations.  For example, the shift
of the Wannier center away from the bond center might serve as a
kind of measure of bond ionicity.  Also, the vector
sum of the Wannier centers immediately gives the bulk electronic
polarization $\bf P$; all three Cartesian components of $\bf
P$ can thus be determined simultaneously using a conventional k-mesh,
instead of constructing separate special k-point strings to compute
each separate Cartesian component of $\bf P$ as is needed
otherwise.\cite{ksv}

But more importantly, the information on the locations of the
Wannier functions may open the possibility of calculating
properties that cannot otherwise be obtained, especially for
distorted, defective, or disordered systems.  For example, it
becomes possible not only to calculate the Born (dynamical)
effective charge $Z^*$, but also to decompose it into displacements
of individual neighboring Wannier centers.  To illustrate this
idea, we have carried out a calculation on a cubic supercell of
GaAs containing 64 atoms ($\Gamma$-only k-point sampling), in which
all atoms are in their equilibrium positions except for 
one Ga atom that is displaced by 0.1 \AA\ along the $[111]$
direction.  Observing the consequent displacement of the Wannier
centers from their bulk crystalline positions, we find a total
$Z_{\rm Ga}^*$ of 2.04, in good agreement with the established
theoretical value of 1.99 as calculated by linear-response
methods.\cite{degir}  Moreover, in arriving at the total electronic
$Z_{\rm Ga}^{*,\rm el}$=-0.96, we find contributions of -1.91,
+0.65, and +0.30 from the groups of four first-neighbor, 12
second-neighbor, and remaining further-neighbor Wannier centers,
respectively.  It is interesting to note that inclusion of
nearest-neighbor contributions alone would thus significantly
overestimate the magnitude of $Z_{\rm Ga}^{*,\rm el}$, and that the
second-neighbor Wannier centers move in the opposite direction to
the Ga atom motion. If we repeat the calculation displacing one As
atom, we obtain a total $Z_{\rm As}^*$ of -2.07 (the acoustic sum
rule\cite{pick} is only approximately satisfied with a finite
k-point sampling). The total electronic $Z_{\rm As}^{*,\rm
el}$=-7.07 has now contributions of -1.74, -4.63, and -0.71 from
the groups of four first-neighbor, 12 second-neighbor, and
remaining further-neighbor Wannier centers, respectively.

In fact, the pattern of displacements of the Wannier centers can
be regarded as defining a kind of coarse-grained representation
of the polarization field, $\bf P(r)$.  To illustrate this idea
more directly, we have carried out a calculation for bulk GaAs
in which a long-wavelength transverse optical (TO) phonon has been
frozen in.  We take the wavevector ${\bf q}=(\pi/4a)(\hat{x}+\hat{y})$ 
($a$ is the lattice parameter) and
relative displacements $\xi({\bf r})=\xi_0\sin({\bf q\cdot r}) \hat{z}$
in a 16-atom supercell, composed of 8 unit cells repeated in the
(110) direction. We assign a displacement amplitude $\xi_0=0.01 a$
to the Ga sublattice, and $-\xi_0$ 
$M_{\rm Ga}/M_{\rm As}$
to the As sublattice ($M_{\rm Ga}$ and $M_{\rm As}$ are the masses 
of the two species; the center of mass doesn't move).
Observing the resulting displacements of the Wannier centers,
we can obtain a picture on how the local polarization changes
from cell to cell (say, by summing all the 4 Wannier centers
surrounding one As atom); fitting these to the same form ${\bf P(r)} = 
P_0 \sin({\bf q\cdot r}) \hat{z}$, we obtain a $P_0$=0.249, and,
via the acoustic sum rule ($Z_{\rm Ga}^{*,\rm el}+Z_{\rm As}^{*,\rm
el}=-8$), we get $Z_{\rm Ga}^{*,\rm el}=-1.52$ and $Z_{\rm
As}^{*,\rm el}=-6.48$.  These results are only in fair agreement
with the bulk values; the discrepancies might be due to the finite
size of our supercell, or to not having used the proper eigenvector
for the phonon mode considered.  However, the main point of this
demonstration is that, given the calculation on the supercell
containing the frozen TO phonon, there is no other way that the
transverse component of the polarization field could have been
obtained.  Since the mode is transverse, $\bf P(r)$ cannot be
determined from the charge density; since ${\bf q}\ne{\bf 0}$, the
Berry-phase approach does not apply; and since the displacement is
finite, the linear-response approach is not directly applicable.
However, the present scheme allows a direct finite-difference
calculation of the transverse polarization field, a quantity that
was previously unavailable.

It would be interesting to apply this kind of analysis to supercell
simulations of amorphous systems such as $a$-H$_2$O or $a$-GaAs.
Once again, while only the longitudinal part of $\bf P(r)$ can be
determined from the charge density, a similar determination of both
the longitudinal and transverse components is possible with access
to the displacements of the Wannier centers, thus leading to a
more complete theory of the dielectric properties of such systems.
This information might be used to assist the approach of
Ref.\ \onlinecite{parritwo}, in which the infrared absorption
spectrum of an amorphous system is extracted from a
molecular-dynamics simulation.
As a limited test, we have carried out calculations
for a 64-atom supercell of crystalline Si with random displacements
typical of $\sim$1000K, and find that the calculation of the
displaced Wannier centers is straightforward.

Finally, we conclude by pointing out that our work opens numerous
possibilities for further development and future study.  On a
practical level, it might be useful to explore the use of more
accurate, higher-order finite difference formulas for $\nabla_{\bf k}$
(see Appendix \ref{app:kmesh}) to see whether convergence
with respect to k-point sampling can be improved.
It might be interesting to apply our analysis within the
semi-empirical tight-binding context, although it should be
noted that matrix elements of $x$, $y$, and $z$ (and, for
$\Omega_{\rm I}$, also of $r^2$) would be needed, in addition to the
Hamiltonian and overlap matrix elements.
Going beyond the scope of the present work,
it might be interesting to other explore localization criteria
e.g., the maximization of the Coulomb self-interaction
of the Wannier functions.  It would also be of great interest to
develop a corresponding theory of maximally-localized {\it
non-orthogonal} Wannier-like functions.  (While the direct
connection to the polarization properties would be lost, there
would be important implications for some linear-scaling
algorithms.)  Finally, there are many questions of a mathematical
character that deserve further study.  For example, is it possible
to prove that our Wannier functions (those that minimize $\Omega$)
have exponential decay, even in the general non-centrosymmetric
multi-band case?  Are they always real, as conjectured in
Sec.\ \ref{sec:real}?  And are there further results that can be
derived regarding the interrelations between the metric tensor, the
Berry connection, and the Berry curvature, as discussed in Appendix
\ref{app:geometry}?  We hope that our work will stimulate some
investigations of these questions.

% -------------------------------------------------------------------

\acknowledgments

This work was supported by NSF grants DMR-96-13648 and ASC-96-25885.
We would like to thank R.~Resta for calling our attention to 
Refs.\ \onlinecite{pati,anandan,joshi};
E.~Stechel for pointing out the connection to
Refs.\ \onlinecite{boys,foster,edmiston,boysr2,weinstein,leonard};
and W.~Kohn, Q.~Niu, and R.~Resta for illuminating
discussions.  

% -------------------------------------------------------------------
\appendix
% -------------------------------------------------------------------

\section{Minimization of spread functional in real space}
\label{app:realmin}

In Sec.\ \ref{sec:rspace} above, the problem of finding the
optimally localized Wannier functions for a periodic system
was formulated directly in real space.  In this Appendix, we
briefly reformulate the problem for the case of a finite system
(cluster, molecule, etc.), and sketch how the minimization of
the functional can be performed in this case.  This provides
a complementary perspective to the k-space procedure discussed
in the main text.

We change notation
$\vert{\bf R}n\rangle\rightarrow\vert i\rangle$ and now refer
to the $i$ as ``localized orbitals'' rather than ``Wannier
functions,'' but their meaning is the same: they are a set of
orthonormal orbitals spanning the Hamiltonian eigenstates in
an energy range of interest (e.g., for the occupied valence
states of a molecule or cluster).

Following the approach of Sec.\ \ref{sec:rspace}, we decompose
$\Omega=\sum_i [\langle r^2\rangle_i-\bar{\bf r}_i^2]$ into an
invariant part $\Omega_{\rm I}=\sum_\alpha{\rm tr}\,[Pr_\alpha
Qr_\alpha]$ (where $P=\sum_i\vert i\rangle\langle i\vert$ and
$Q=1-P$) and a remainder
$\widetilde\Omega = \sum_\alpha \sum_{i\ne j} \, \vert\langle i
\vert r_\alpha \vert j \rangle \vert^2$.
Defining matrices $X_{ij}=\langle i \vert x \vert j \rangle$,
$X_{{\rm D},ij}=X_{ij}\,\delta_{ij}$, $X'=X-X_{\rm D}$, and
similarly for $Y$ and $Z$, this can be rewritten
\begin{equation}
\widetilde\Omega = {\rm tr}\,[X'^2+Y'^2+Z'^2] \;\;.
\label{eq:otr2}
\end{equation}
Thus if $X$, $Y$, and $Z$ could be simultaneously diagonalized, then
$\widetilde\Omega$ could be minimized to zero, but for non-commuting
matrices this is not possible.  In a sense, our job is to perform
the optimal approximate simultaneous co-diagonalization of the three
Hermitian matrices $X$, $Y$, and $Z$ by a single unitary
transformation.  We are not aware of a formal solution for this
problem, but a steepest-descent numerical solution is fairly
straightforward.  Since ${\rm tr}\,[X'X_{\rm D}]=0$, etc.,
\begin{equation}
d\Omega = 2\,{\rm tr}\,[X'dX+Y'dY+Z'dZ] \;\;.
\label{eq:dotr}
\end{equation}
We consider an infinitesimal unitary transformation $\vert
i\rangle \;\rightarrow\; \vert i\rangle +\sum_j W_{ji} \vert
j\rangle$ (where $dW$ is antihermitian),
from which $dX=[X,dW]$, etc.  Inserting in Eq.\ (\ref{eq:dotr})
and using ${\rm tr}\,[A[B,C]]=\,{\rm tr}\,[C[A,B]]$ and
$[X',X]=[X',X_{\rm D}]$, we obtain
$d\Omega={\rm tr}\,[dW\,G]$ where
\begin{equation}
G = 2\,\Bigl\{ [X',X_{\rm D}] + [Y',Y_{\rm D}]
+ [Z',Z_{\rm D}] \Bigr\} \;\;,
\label{eq:Greal}
\end{equation}
so that the desired gradient is ${d\Omega/dW} = G$ as given
above.  The minimization can then be
carried out using steepest descents following the general approach
outlined in Sec.\ \ref{sec:sd}.  More sophisticated but related
methods are discussed in Ref.\ \onlinecite{leonard}.

If this approach is applied to a finite system having a
crystalline interior, the solutions in the interior are expected
to correspond precisely with the maximally-localized Wannier functions
as determined using the k-space methods of the main text.
In the vicinity of surfaces or defects, or for disordered materials,
the solutions will essentially correspond to the ``generalized
Wannier functions'' discussed by previous
authors.\cite{onffroy,rehr,geller,nenciu93}

% -------------------------------------------------------------------

\section{Finite-difference formulas for \lowercase{k}-space grids}
\label{app:kmesh}

We assume that the Brillouin zone has been discretized into a
uniform Monkhorst-Pack mesh.\cite{monkpack} Let $\bf b$ be a
vector connecting a k-point to one of its near neighbors, and
let $Z$ be the number of such neighbors to be included in the
finite-difference formulas.  We seek the simplest possible
finite-difference formula for $\nabla_{\bf k}$, i.e., the one
involving the smallest possible $Z$.
When the Bravais lattice point group
is cubic, it will only be necessary to include the first shell
of $Z=6$, 8, or 12 k-neighbors for simple cubic, bcc, or
fcc k-space meshes, respectively.  Otherwise, further shells
must be included until it is possible to satisfy the condition
\begin{equation}
\sum_{\bf b} w_b \, b_\alpha b_\beta = \delta_{\alpha\beta}
\label{eq:condit}
\end{equation}
by an appropriate choice of a weight $w_b$ associated with each
shell $|{\bf b}|=b$.  For the three kinds of cubic mesh,
Eq.\ (\ref{eq:condit}) is satisfied with $w_b=3/Zb^2$ (single
shell).  Taking next the slightly more complicated case of an
orthorhombic lattice, one can let $\bf b$ run over the two nearest
neighbors in each Cartesian direction ($Z=6$), with $w_b=1/2b_x^2$
for the two neighbors at $\pm b_x\hat{x}$, etc.  Even in the worst
case of minimal (triclinic) symmetry, only six pairs of neighbors
($Z=12$) should be needed, as the freedom to choose six weights
should allow one to satisfy the six independent conditions comprising
Eq.\ (\ref{eq:condit}).

Now, if $f(\bf k)$ is a smooth function of $\bf k$, its gradient can
be expressed as
\begin{equation}
\nabla f({\bf k})=\sum_{\bf b} w_b \, {\bf b}
 \,[f({\bf k+b})-f({\bf k})] \;\;.
\label{eq:grad}
\end{equation}
We can check the correctness of this finite-difference formula
by applying it to the case of a linear function
$f({\bf k})=f_0+{\bf g\cdot k}$, for which we find
$\nabla_\alpha f({\bf k})= \sum_{\bf b} w_b 
\sum_\beta b_\alpha g_\beta b_\beta =g_\alpha$.  In a similar way,
\begin{equation}
\vert\nabla f({\bf k})\vert^2=\sum_{\bf b} w_b
 \,[f({\bf k+b})-f({\bf k})]^2 \;\;.
\label{eq:nabla}
\end{equation}

We note that improved accuracy and k-set convergence might be
obtained by utilizing improved, higher-order finite-difference
formulas involving more shells of neighboring k-points, but we have
not explored this possibility here.

% -------------------------------------------------------------------

\section{Geometric properties and complexity of electron bands}
\label{app:geometry}

Consider a manifold of $J$ orthonormal states
$\vert\psi_n(\lambda)\rangle$, $n=1,...,J$, depending on a
continuous $d$-dimensional parameter $\lambda$.  Alternatively,
one can view these as representing the projection
$P(\lambda)=\sum_n \vert\psi_n(\lambda)\rangle
\langle\psi_n(\lambda)\vert$.  For the application to electron
bands in crystals, we identify $\lambda\rightarrow\bf k$ and
$\psi_n(\lambda)\rightarrow u_{n\bf k}$.  Here, we investigate
the geometric properties of such a manifold, generalizing the
single-state ($J=1$) results of 
Refs.\ \onlinecite{pati,anandan,joshi} to the multi-state case.

One can define two kinds of intrinsic geometric properties: a
{\it geometric distance} and a {\it geometric phase}.  We
consider the former first.  The geometric distance $D_{12}$
between two points $\lambda_1$ and $\lambda_2$ is here taken to be
\begin{equation}
D_{12}^2\,=\,{\rm tr}[P_1Q_2]\,=\,{\textstyle{1\over 2}}\,
\Vert P_1-P_2 \Vert^2 \;\;,
\label{eq:dist}
\end{equation}
where $Q(\lambda)=1-P(\lambda)$.  In the case of a single state,
this becomes $D_{12}^2=1-|\langle\psi_1\vert\psi_2\rangle|^2$,
which for small separations is consistent with the slightly different
definition $D_{12}^2=2-2|\langle\psi_1\vert\psi_2\rangle|$ of
Ref.\ \onlinecite{pati}.  Considering the distance for infinitesimal
separations, one can define a Riemannian metric\cite{pati}
\begin{equation}
D_{\lambda,\lambda+d\lambda}^2\,=\sum_{\alpha\beta} \, g_{\alpha\beta}
\, d\lambda_\alpha \, d\lambda_\beta \;\;.
\label{eq:Ddef}
\end{equation}
Introducing the notation $\psi_{n,\alpha} = d\psi_n/d\lambda_\alpha$,
etc., and making use of
\begin{equation}
0=\langle\psi_n|\psi_{m,\alpha}\rangle+\langle\psi_{n,\alpha}|
   \psi_m\rangle \;\;,
\label{eq:forder}
\end{equation}
\begin{equation}
0=\langle\psi_n|\psi_{m,\alpha\beta}\rangle
 +\langle\psi_{n,\alpha\beta}|\psi_m\rangle
 +2\,{\rm Re}\,\langle\psi_{n,\alpha}|\psi_{m,\beta}\rangle
\;\;,
\label{eq:sorder}
\end{equation}
which follow from the fact that the $\psi_n$ remain orthonormal at
first and second order in $d\lambda$, the metric $g_{\alpha\beta}$
becomes, after some manipulation,
\begin{equation}
g_{\alpha\beta}=
   \,{\rm Re}\,\sum_n\,\langle\psi_{n,\alpha}|\psi_{n,\beta}\rangle
   -\,\sum_{mn}\,\langle\psi_{n,\alpha}|\psi_m\rangle
               \,\langle\psi_m|\psi_{n,\beta}\rangle \;\;,
\label{eq:gab}
\end{equation}
which reduces in the single-band case to the expression of
Pati.\cite{pati}

From Eq.\ (\ref{eq:dist}) it is obvious that the distance, and thus
the metric, are gauge-invariant quantities.  These are therefore
intrinsic properties of the manifold.  One way of thinking about
the metric is to observe that for any given path in $\lambda$ space,
the line integral of $g^{1/2}$ along the path provides a measure
of the total ``quantum distance'' along the path; intuitively,
it is a measure of the amount of change of character of the states
as one traverses the path.  The physical meaning of this distance for
the case of temporal evolution of quantum states is discussed in
Refs.\ \onlinecite{pati,anandan,joshi}.

The second type of geometric object that can be defined is a
``geometric phase'' or ``Berry phase.''\cite{berry}  Here, one is
interested in considering closed paths in $\lambda$ space, and
relating the phase (or, for the multi-state case, the unitary
rotation) induced by adiabatic (``parallel'') transport along the
path.  The multi-state (``non-Abelian'') case has been discussed by
Wilczek and Zee\cite{wilczek}, Mead,\cite{mead} and
Resta.\cite{resta96} One can define a (non-gauge-invariant) Berry
connection
\begin{equation}
A_{\alpha,nm}=i\,\langle\psi_n|\psi_{m,\alpha}\rangle
\end{equation}
and a (gauge-covariant) Berry curvature
\begin{equation}
B_{\alpha\beta}^{nm}=-\partial_\alpha A_{\beta,nm}
    +\partial_\beta A_{\alpha,nm} +i\,[A_\alpha,A_\beta]_{nm} \;\;.
\end{equation}
The invariants of the latter, such as
\begin{equation}
{\rm tr}\,B_{\alpha\beta} = 2\,{\rm Im}\,\sum_n
   \, \langle\psi_{n,\alpha}|\psi_{n,\beta}\rangle \;\;,
\label{eq:trY}
\end{equation}
[see Eq.\ (3.29) of Ref.\ \onlinecite{resta96}]
are thus gauge-invariant.  (We shall use the notation `tr' and
`Tr' to denote electronic and Cartesian traces, respectively.)

There is a tantalizing similarity between the metric $g_{\alpha\beta}$,
Eq.\ (\ref{eq:gab}), and the quantum trace of the Berry curvature,
Eq.\ (\ref{eq:trY}).  In fact, defining the gauge-invariant
quantity
\begin{equation}
{\cal F}_{\alpha\beta} = \sum_n \, \langle\psi_{n,\alpha}
|Q|\psi_{n,\beta}\rangle
\end{equation}
where again $Q=1-P$,
and using Eq.~(\ref{eq:forder}) to show that the second term
in Eq.\ (\ref{eq:gab}) is intrinsically real, we obtain simply
$g_{\alpha\beta}={\rm Re}\,{\cal F}_{\alpha\beta}$ and
${\rm tr}\,B_{\alpha\beta}=
2\,{\rm Im}\,{\cal F}_{\alpha\beta}$.
This suggests that there may be some deep connections between
the two quantities.\cite{pati,anandan,joshi}  In the case
where the states $\psi_n$ are eigenstates of a Hamiltonian
$H(\lambda)$, one moreover has\cite{resta96}
\begin{equation}
{\cal F}_{\alpha\beta} = \sum_{n=1}^J \, \sum_{m=J+1}^\infty\;
{  \langle\psi_n|H_\alpha|\psi_m\rangle
   \langle\psi_m|H_\beta |\psi_n\rangle
\over (E_n-E_m)^2 }
\end{equation}
where $H_\alpha=dH(\lambda)/d\lambda_\alpha$.

We now return to the case of electron bands in
crystals, $\lambda\rightarrow\bf k$ and $\psi_n(\lambda)
\rightarrow u_{n\bf k}$, and discuss the geometric properties
induced by the band projection operator $P^{\bf(k)}$.
Note that $g$, $A$, and $B$ have units of $l^2$, $l$, and
$l^2$, respectively.  Again focusing first on the metric, and
comparing Eq.\ (\ref{eq:omI}) with the definitions (\ref{eq:dist})
and (\ref{eq:Ddef}), we find
\begin{equation}
\Omega_{\rm I}\,=\,{1\over N}\,\sum_{\bf k,b}\sum_{\alpha\beta}\,
w_b\,g_{\alpha\beta} \,b_\alpha\,b_\beta
\end{equation}
or, using Eq.\ (\ref{eq:condit}) and restoring the continuum limit,
\begin{equation}
\Omega_{\rm I} \,=\,{V\over (2\pi)^3}\,\int_{\rm BZ} d{\bf k} \,
{\rm Tr}\,g{(\bf k)} \;\;,
\end{equation}
where the integral is over the Brillouin zone.  Thus, the invariant
part of the spread functional is nothing other than the
Brillouin-zone average of the trace of the metric!

It may be interesting to see whether other global properties
of the metric might be given some physical interpretation.
In particular, we define a dimensionless and gauge-invariant
quantity
\begin{equation}
C=\int_{\rm BZ} d{\bf k}\;{\rm det}^{1/2}\,g({\bf k}) \;\;.
\label{eq:complexity}
\end{equation}
We shall call this the ``complexity'' of the bands.  Mathematically,
it is really nothing other than the volume of the
Brillouin zone as measured according to the metric $g$.  However,
we have called it the ``complexity'' because it measures the
variation of the character of the band projection operator $P^{(\bf
k)}$ throughout the Brillouin zone.  Everything said here applies to
any isolated band or composite group of bands, but we have in mind
primarily the case where all the occupied valence bands in an
insulator are considered as a composite group.  In this case, and
assuming that one is only interested in quantities (such as total
energies and forces) that can be expressed as a trace over the
bands, the complexity might thus be expected to reflect (and even
predict) the number of k-points needed for an accurate sampling of
the Brillouin zone.  We have not tested this idea numerically, but
this would clearly be an interesting avenue for future exploration.

Turning now to phase properties, we note that a finite-different
representation of the Berry connection is
\begin{equation}
A_{\alpha,mn}\,=\,i\,\sum_b\,w_b\,b_\alpha
\left[ \, M_{mn}^{\bf(k,b)}\,-\,\delta_{mn}\,\right] \;\;.
\label{eq:bcfdiff}
\end{equation}
Restoring the continuum limit in $k$-space, we can write
\begin{equation}
\bar{\bf r}_n\,=\, {V\over (2\pi)^3}\,\int_{\rm BZ} d{\bf k} \,
{\bf A}_{nn}({\bf k})\;\;,
\label{eq:bcr}
\end{equation}
and more generally,
\begin{equation}
\langle{\bf 0}m|{\bf r}|{\bf R}n\rangle\,=\,
  {V\over (2\pi)^3} \,\int_{\rm BZ} d{\bf k} \, {\bf A}_{mn}({\bf k})
  \,e^{i\bf k\cdot R} \;\;.
\label{eq:AofR}
\end{equation}
The right-hand side is just ${\bf A}_{mn}({\bf R})$, the Fourier
coefficient of the Berry curvature.
Eq.\ (\ref{eq:bcr}) is just the expression for the position of
the Wannier center, which contributes to the electronic
polarization.\cite{blount,ksv,resta,resta96}
Moreover,
\begin{equation}
\widetilde{\Omega}_{\rm D}\,=\, \sum_n\,
{V\over (2\pi)^3}\,\int_{\rm BZ} d{\bf k} \,
\left\vert\,{\bf A}_{nn}({\bf k})\,-\,\bar{\bf r}_n
\,\right\vert^2 \;\;,
\label{eq:bcod}
\end{equation}
\begin{equation}
\widetilde{\Omega}_{\rm OD}\,=\, \sum_{m\ne n}\,
{V\over (2\pi)^3}\,\int_{\rm BZ} d{\bf k} \,
\left\vert\,{\bf A}_{mn}({\bf k})
\,\right\vert^2 \;\;.
\label{eq:bcood}
\end{equation}
Eqs.\ (\ref{eq:bcod}-\ref{eq:bcood}) show that the non-invariant
parts of the spread functional are also conveniently written in terms
of the Berry connection.  If the above equations are reexpressed
in terms of the Fourier coefficients $A_{mn}\bf(R)$,
Eqs.\ (\ref{eq:omod}) and (\ref{eq:omd}) are immediately recovered.

In the single-band case, we showed in Sec.\ \ref{sec:isoband}
that the minimum value of $\widetilde{\Omega}$ could be related
to the transverse part of the Berry connection, which in turn is
determined by the gauge-invariant Berry curvature.
In the multiband case, the Berry curvature
$B_{\alpha\beta}^{mn}({\bf k})$ is no longer gauge-invariant, and
it is not obvious whether it is possible to make a corresponding
decomposition.  Nevertheless, one can derive similar correspondences
as those above for $\bf A$.  So,
\begin{equation}
B_{\alpha\beta}^{mn}({\bf k})\,=\,
  -i\,\langle\,u_{m,\alpha}\,|\,Q\,|\,u_{n,\beta}\,\rangle\,
  +i\,\langle\,u_{m,\beta}\,|\,Q\,|\,u_{n,\alpha}\,\rangle\, \;\;,
\end{equation}
\begin{equation}
B_{\alpha\beta}^{mn}({\bf R})\,=\,
  -i\,\langle\,u_m\,\vert\,r_\alpha Qr_\beta \,-\,
  r_\beta Qr_\alpha\,\vert\,u_n\,\rangle \;\;.
\end{equation}
Making use of
$r_\alpha Qr_\beta-r_\beta Qr_\alpha=[Pr_\alpha P,Pr_\beta P]$,
one finds
\begin{eqnarray}
\Vert\,[Pr_\alpha P,Pr_\beta P]\,\Vert_{\rm c}^2
\,&=&\,
\sum_{\bf R}\,\sum_{mn}\,|\,B_{\alpha\beta}^{mn}({\bf R})\,|^2
\nonumber \\
\,&=&\, {V\over{(2\pi)^3}}\,\int_{\rm BZ}\,d{\bf k}\, \left\Vert\,
B_{\alpha\beta}({\bf k})\,\right\Vert^2 \;\;.
\label{eq:trbcurv}
\end{eqnarray}
Each form above is manifestly gauge-invariant and positive-definite.
Thus, it can be seen that the Berry curvature will vanish if and
only if the band-projected position operators $PxP$, $PyP$, and $PzP$
commute with one another; as discussed following
Eq.\ (\ref{eq:eigfun}), this is also just the condition that
$\widetilde{\Omega}$ vanishes at the minimum.

% -------------------------------------------------------------------

% -------------------------------------------------------------------


\begin{references}
\bibitem{expl:phase} We find empirically that the
maximally localized Wannier functions are always real, once the
arbitrary overall phase is removed.  This observation is discussed
and justified in Sec.\ \ref{sec:real}.
\end{references}
\end{document}